\documentclass{aa}
\usepackage{txfonts}
\usepackage{graphicx}
\usepackage{natbib}
\usepackage{epstopdf}

\begin{document}

\title{The nearby population of M-dwarfs with WISE: A search for warm circumstellar dust}

\author{Henning Avenhaus \and Hans Martin Schmid \and Michael R. Meyer}

\institute{Institute for Astronomy, ETH Zurich, Wolfgang-Pauli-Strasse 27, CH-8093 Zurich, Switzerland}

\date{Received 11 June 2012 / Accepted 30 August 2012}

\abstract{Circumstellar debris disks are important because of their connection and interaction with planetary systems. An efficient way to identify these systems is through their infrared excess. Most studies so far concentrated on early-type or solar-type stars, but less effort has gone into investigating M-dwarfs, which pose the additional problem that their mid-infrared colors are poorly known.}
{We characterize the infrared photometric behavior of M-dwarfs and search for infrared excess in nearby M-dwarfs taken from the volume-limited RECONS sample, concentrating on mid-infrared wavelengths corresponding to warm ($\gtrsim$ 100 K) dust. We then check whether the population of these late-type stars has a significantly different fraction of infrared excess compared to earlier-type stars.}
{We used data recently released from the \emph{WISE} satellite, which provides the most sensitive mid-infrared all-sky survey to date. Our sample consists of 85 sources encompassing 103 M-dwarfs. We compared this sample to \emph{Spitzer} data and matched it to the 2MASS catalog. We derived empirical infrared colors from these data and discuss the intrinsic spread within these colors as well as the errors from \emph{WISE} and 2MASS. Based on this, we checked the stars for infrared excess and discuss the minimum excess we would be able to detect.}
{Other than the M8.5 dwarf SCR 1845-6357 A, where the mid-infrared excess is produced by a known T6 companion, we detect no excesses in any of our sample stars. The limits we derive for the 22$\mu$m excess are slightly higher than the usual detection limit of $\sim$10-15$\%$ for Spitzer studies, but including the 12$\mu$m band and the $[12]-[22]$ color in our analysis allows us to derive tight constraints for the fractional dust luminosity $L_{dust}/L_{\star}$. Comparing our findings to earlier-type stars, we show that this result is consistent with M-dwarf excesses in the mid-inrared being as frequent as excesses around earlier-type stars. The low detection rate of $0^{+1.3}_{-0.0}\%$ we derive for our sample in that case is an age effect. We also present a tentative excess detection at 22$\mu$m around the known cold debris disk M-dwarf AU Mic, which is not part of our statistical sample.}
{There is still no clear detection of a mid-infrared excess around any old ($\gtrsim$30 Myr) main-sequence M-dwarf. It is unclear whether this is due to a different dust evolution around M-dwarfs compared to earlier-type stars or whether this is an age effect combined with the difficulties involved in searching M-dwarfs for infrared excesses. A significantly larger sample of well-studied M-dwarfs is required to solve this question. At the same time, their behavior at longer wavelengths, which are sensitive to colder dust, needs further investigation.}

\titlerunning{M-dwarfs with WISE: Mid-infrared properties and excesses}
\authorrunning{Avenhaus et al.}

\maketitle

\section{Introduction}

Dust around main-sequence stellar systems has become a very well-studied phenomenon since it was first seen. The first detection was made around Vega, an A0 star, through the detection of infrared excess \citep{Aumann1984}. A lot of stars showing this "Vega phenonenon" were subsequently detected and imaged through the all-sky survey performed by \emph{IRAS} at 25, 60 and 100 $\mu$m \citep{Zuckerman2001, Rhee2007}, as well as with pointed observations using \emph{ISO} \citep{Decin2003, Spangler2001} at 60 and 100 $\mu$m, \emph{Spitzer} \citep{Rebull2008, Carpenter2009, Koerner2010} at 24 and 70 $\mu$m, and \emph{Herschel} (most recently by \citealp{Acke2012}). Recently, the all-sky \emph{AKARI} satellite was able to make detections at 9 and 18$\mu$m wavelength \citep{Fujiwara2009}.

It was quickly realized that the infrared excess must stem from dust that surrounds the star in a disk. It was furthermore understood that micron-sized dust cannot persist in orbit around a star for very long due to effects such as radiation pressure, Poynting-Robertson drag, and stellar wind pressure. Thus, the dust has to be permanently regenerated in situ. The standard interpretation of this is that the debris traces underlying planetesimal belts, where dust is generated through collisions \citep{Backman1993}. The same effect happens in the solar system's asteroid belt and is responsible for zodiacal light. The difference to the solar system is that these belts are mostly collisionally dominated, producing a much stronger infrared excess signal than the solar system debris disk does.

The first detection of this phenomenon was made around an A-star. Because these bright stars are easiest to observe, the detection of an infrared excess is relatively easy for A, F, G, and nearby K-stars. Late-type star and specifically M-dwarf infrared excesses are significantly harder to detect. First, they are much fainter, and secondly, the Rayleigh-Jeans approximation does not hold in the mid-infrared for M-dwarfs. Molecular bands dominate their spectra at visible and near-infrared wavelengths. Because of this, the expected scatter in their mid- and far-infrared colors is larger compared to early-type stars. Most searches for infrared excess concentrate on early-type or solar-type stars (e.g., \citealp{Su2006, Bryden2009, Moor2011}).

A second effect is that the frequency of debris disks declines with age \citep{Carpenter2009}. Vega has an age of only $\sim$450 Myr \citep{Yoon2010}, and many studies for infrared excesses around early-type stars include very young targets (e.g. \citealp{Su2006}). Very few young main-sequence M-dwarfs are known; one of them is the well-studied debris disk system AU Mic (e.g., \citealp{Fitzgerald2007}). This is because of the intrinsic faintness of M-dwarfs, which only allows detecting them if they are close to the solar system. M-dwarf debris disk surveys are therefore usually bound to target mostly old stars, where the excess frequency is expected to be lower. Of course, many TTauri stars are M-dwarfs with strong mid-IR excess from primordial disk emission and there is also growing evidence that primordial disks last longer around lower-mass stars \citep{Carpenter2006}, but no mid-infrared excesses attributable to debris have been found except around AT Mic, like AU Mic a member of the $\sim$ 12 Myr $\beta$ Pictoris moving group \citep{Gautier2007, Plavchan2009}. Few detections of cold, Kuiper-belt-like debris disks have been made in the sub-mm regime for young M-dwarfs \citep{Liu2004, Lestrade2009}, at a seemingly lower rate than for earlier-type stars. It has been pointed out that there is an apparent lack of debris disks around older M-dwarfs in general \citep{Plavchan2005}.

On the other hand, M-dwarfs make up the majority of stars in our galaxy, which is a good reason to study them. Because of its magnitude limits and the intrinsic faintness of M-dwarfs, \emph{IRAS} can only give access to the very closest M-dwarf systems. The recently flown \emph{WISE} satellite does not have the far-infrared capabilities of \emph{IRAS}, but goes significantly deeper in the 12- and 22-$\mu$m bands, giving access to more M-dwarf systems \citep{Wright2010}. M-dwarf studies going this deep have been possible earlier only with dedicated pointed observations, for example using the \emph{Spitzer} space telescope \citep{Werner2004}.

Theoretical studies of the behavior and evolution of dust around M-dwarfs suggest that the timescales and frequencies involved in M-dwarfs might be considerably different from what we observe in earlier-type stars \citep{Plavchan2005}. Observationally, there is a lack of solid evidence for a difference in either the frequency or the timescales to disperse debris disks when comparing M-dwarfs to earlier-type stars. No studies were able to prove a statistically significant difference between A-stars and solar-type stars, either \citep{Trilling2008}. The question whether the processes involved in the evolution of debris disks work similar over a wide range of stellar luminosities or whether there are significant differences is so far unresolved.

We describe the data we acquired from \emph{WISE} and \emph{2MASS} in section 2. In section 3, we describe the analysis of the data, our efforts to detect mid-infrared excesses and the limits we can derive from this on the excess at different wavelengths and on the dust surrounding the M-dwarfs in our sample. In section 4, we compare our study to other results from other M-dwarf surveys as well as surveys of earlier-type stars. We discuss implications in section 5 and draw our conclusions in section 6.

\section{Source data from RECONS, WISE, and 2MASS}

The \emph{WISE} satellite was launched on December 14, 2009 and began its operations in January 2010, scanning the complete sky in four wavebands centered at 3.4, 4.6, 12, and 22 $\mu$m, named w1 through w4 \citep{Wright2010}. On April 14, 2011, a preliminary data release covered about 57 percent of the sky. On March 14, 2012, the all-sky catalog was released. The \emph{WISE} satellite is significantly more sensitive than the \emph{IRAS} satellite was, and is also more sensitive than the more recent \emph{AKARI} sky survey \citep{Murakami2007}. Its 5-$\sigma$ detection limits are 16.4, 15.4, 11.1, and 7.8 magnitudes, respectively. The \emph{WISE} beams have a FWHM of 6.1$''$, 6.4$''$, 6.5$''$, and 12$''$, respectively. This means that close binary systems cannot be resolved, an aspect we will have to take into account.

The 2MASS survey covered the entire sky in the $J$, $H$, and $K_s$ wavebands during the years 1997 through 2001 using ground-based facilities \citep{Skrutskie2006}. Its sensitivity as well as its resolution are better than the \emph{WISE} limits. The errors in the 2MASS magnitudes stem mainly from atmospheric effects and calibration uncertainties.

\subsection{Stellar sample}

Because of the intrinsic faintness of late-type dwarfs, and because of the magnitude limit of \emph{WISE}, which is still relatively strict at 7.8 magnitudes in the longest waveband, a study of the mid-infrared behavior and excesses has to focus on nearby M-dwarfs. Very late M-dwarfs (M7, M8, M9) are too faint for the \emph{WISE} 22 $\mu$m band even at distances of only 10 pc, while early-type M-dwarfs (M0, M1) can be usefully tracked out to distances of 25 pc or more. Because our aim is to have a sample not biased heavily toward early-type M-dwarfs, and because M8 and M9 dwarfs are intrinsically rarer than M0 and M1 dwarfs (the IMF peaks around M3-M4, \citealp{Bastian2010}), we concentrate on the immediate solar neighborhood, omitting many early-type M-dwarfs that would be within reach of \emph{WISE}, but are beyond 10pc.

One of the best and most complete samples of the immediate solar vicinity is the list of the 100 nearest stellar systems by the RECONS team \citep{Jao2005, Henry2006}. Including eight systems that were pushed out of the top 100 list, it consists of 108 systems encompassing 139 stars (not counting planets and white or brown dwarfs, which also appear in the catalog). The vast majority of these stars are M-dwarfs (110). One of the advantages of the RECONS sample is that in addition to giving accurate positions and proper motions (required because these very near objects move significantly on sky between the 2MASS and \emph{WISE} surveys), it also gives accurate parallaxes, which can be used to determine absolute luminosities. V magnitudes are also provided. RECONS can thus give a clean, very complete volume-limited sample, which is ideal for our purpose. It was also readily available at the time we started our analysis, in contrast to, for example, the Reid 8pc sample \citep{Reid2007}, which would have been another option. The number of M-dwarfs in the Reid sample is slightly larger. However, the statistical analysis we aimed to perform can be performed equally well with the RECONS sample, and the slightly larger number of M-dwarfs in the Reid sample would not change our discussion significantly.

\subsection{Observational data}

\begin{table}
\caption{Excluded stars}
\label{table:excludedstars}    
\centering
\begin{tabular}{lp{5.5cm}}
\hline\hline\noalign{\smallskip}
{Name} &     {reason for exclusion}\\
\noalign{\smallskip}
\hline
\noalign{\smallskip}
{GJ 783 B}&{confusion with nearby (7$"$) K2.5 primary}\\
{GJ 338 A}&{confusion with nearby (17$"$) K7 secondary}\\
{LP 771-095 A/B/C}&{WISE catalog shows multiple sources at expected position, but none of them matches the star}\\
{GJ 725 A/B}&{close binary system (14$"$) that has very bad S/N ratio due to mutual contamination}\\
{GJ 752 B}&{excluded from analysis of w4 band, not detected in WISE w4}\\
\noalign{\smallskip}
\hline
\end{tabular}
\tablefoot{Stars that were excluded from the analysis for reasons of data quality.}
\end{table}

\begin{figure}
\includegraphics[width=9cm]{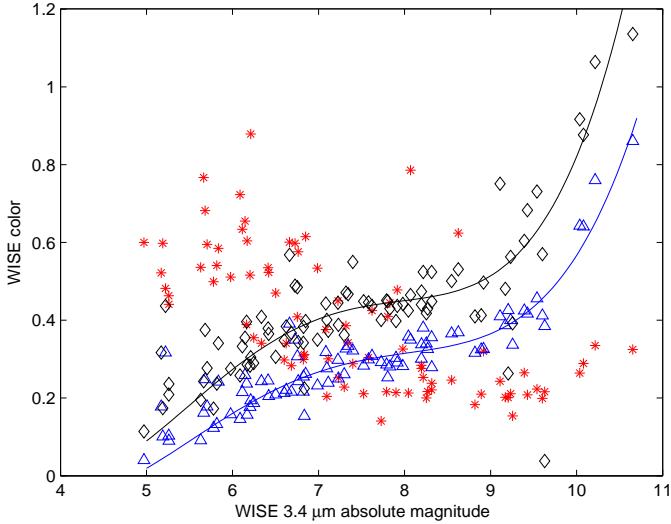}
\caption{WISE $[3.4]$-$[4.6]$ color (red stars), $[3.4]$-$[12]$ color (black diamonds) and $[3.4]$-$[22]$ color (blue triangles), plotted against the 3.4 $\mu$m absolute magnitude. Both the $[3.4]$-$[12]$ and the $[3.4]$-$[22]$ color show the expected trend (they become redder for late-type stars) and can be fitted with a polynomial. The $[3.4]$-$[4.6]$ color is expected to show a similar trend and furthermore be smaller than the other two colors for any spectral type. This is clearly not the case for the bright end of our sample (above magnitudes of $\sim$6 or absolute magnitudes of $\sim$8-9). We therefore omit the second channel in our analysis.}
\label{fig:SecondChannelBad}
\end{figure}

All M-dwarfs in the RECONS sample can be seen at the expected positions in the \emph{WISE} images. Because some of the stars are confused with nearby earlier stars and/or tight binary M-dwarf companions, there are however only 90 individual sources for the 110 M-dwarfs in the catalog. Out of these, five sources containing seven stars (GJ 783 B, GJ 338 A, LP 771-095 A/B/C,
GJ 725 A and GJ 725 B) had to be removed from our analysis because of poor data quality, mostly from confusion with nearby sources. One more star (GJ 752 B) had to be excluded from the analysis of the 22 $\mu$m band because it could not be detected at this wavelength. A summary of the excluded sources and the reasons for exclusion is presented in Table \ref{table:excludedstars}. This leaves 103 M-dwarfs in 85 individual \emph{WISE} sources (70 single, 12 double and 3 triple systems) for our analysis (one single star less for the w4 band).

\begin{table*}[ht!]
\caption{Observational data}
\label{table:observationaldata}    
\centering
\begin{tabular}{lcr@{$\pm$}lccr@{.}lr@{$\pm$}lr@{$\pm$}lr@{$\pm$}lr@{$\pm$}lr@{$\pm$}lr@{$\pm$}l} 
\hline\hline\noalign{\smallskip}
{}&        {stars in}&        \multicolumn{2}{c}{Distance}&	{Spectral}&	{SpTy}&    \multicolumn{2}{c}{V}&        \multicolumn{2}{c}{K$_s$}&        \multicolumn{2}{c}{$F_{3.4}$}&        \multicolumn{2}{c}{$F_{12}$}&        \multicolumn{2}{c}{$F_{22}$} \\
{Name} &  {beam} & \multicolumn{2}{c}{(pc)$^{a}$}& 	{Type$^{b}$}&	{ref.}&           \multicolumn{2}{c}{(mag)$^{c}$}&    \multicolumn{2}{c}{(mag)$^{d}$}&    \multicolumn{2}{c}{(mag)$^{e}$}&    \multicolumn{2}{c}{(mag)$^{e}$}&    \multicolumn{2}{c}{(mag)$^{e}$}\\

\noalign{\smallskip}
\hline
\noalign{\smallskip}
{GJ 551}&	{1}&	{1.30}&{0.01}&	{M5.5}&	{1}&	{11}&{05}&	{4.38}&{0.03}&	{4.20}&{0.09}&	{3.83}&{0.02}&	{3.66}&{0.02$^{f}$}\\{GJ 699}&	{1}&	{1.83}&{0.01}&	{M4.0}&	{1}&	{9}&{57}&	{4.52}&{0.02}&	{4.39}&{0.07}&	{4.04}&{0.02}&	{3.92}&{0.03$^{f}$}\\{GJ 406}&	{1}&	{2.39}&{0.01}&	{M6.0}&	{2}&	{13}&{53}&	{6.08}&{0.02}&	{5.81}&{0.06}&	{5.48}&{0.02}&	{5.31}&{0.03$^{f}$}\\{GJ 411}&	{1}&	{2.54}&{0.01}&	{M2.0}&	{1}&	{7}&{47}&	{3.25}&{0.31}&	{3.24}&{0.14}&	{3.05}&{0.01}&	{2.93}&{0.02$^{f}$}\\{GJ 65 A/B}&	{2}&	{2.68}&{0.02}&	{M5.5 / M6.0}&	{1}&	{12}&{06}&	{5.34}&{0.02}&	{5.05}&{0.07}&	{4.76}&{0.02}&	{4.62}&{0.03}\\{GJ 729}&	{1}&	{2.97}&{0.02}&	{M3.5}&	{1}&	{10}&{44}&	{5.37}&{0.02}&	{5.16}&{0.06}&	{4.91}&{0.01}&	{4.72}&{0.03$^{f}$}\\{GJ 905}&	{1}&	{3.16}&{0.01}&	{M5.5}&	{1}&	{12}&{29}&	{5.93}&{0.02}&	{5.69}&{0.06}&	{5.39}&{0.02}&	{5.25}&{0.03$^{f}$}\\{GJ 887}&	{1}&	{3.28}&{0.01}&	{M0.5}&	{3}&	{7}&{34}&	{3.47}&{0.20}&	{3.24}&{0.12}&	{3.08}&{0.01}&	{3.00}&{0.02$^{f}$}\\{GJ 447}&	{1}&	{3.35}&{0.02}&	{M4.0}&	{1}&	{11}&{16}&	{5.65}&{0.02}&	{5.46}&{0.06}&	{5.18}&{0.01}&	{5.03}&{0.03$^{f}$}\\{GJ 866 A/B/C}&	{3}&	{3.45}&{0.05}&	{M5.0}&	{1}&	{12}&{30}&	{5.54}&{0.02}&	{5.31}&{0.06}&	{5.01}&{0.02}&	{4.88}&{0.03}\\{GJ 15 A}&	{1}&	{3.57}&{0.01}&	{M1.5}&	{1}&	{8}&{08}&	{4.02}&{0.02}&	{3.85}&{0.10}&	{3.71}&{0.02}&	{3.60}&{0.02$^{f}$}\\{GJ 15 B}&	{1}&	{3.57}&{0.01}&	{M3.5}&	{1}&	{11}&{06}&	{5.95}&{0.02}&	{5.75}&{0.05}&	{5.46}&{0.02}&	{5.30}&{0.03$^{f}$}\\{GJ 1111}&	{1}&	{3.63}&{0.04}&	{M6.5}&	{1}&	{14}&{90}&	{7.26}&{0.02}&	{7.03}&{0.03}&	{6.63}&{0.02}&	{6.47}&{0.06$^{f}$}\\{GJ 1061}&	{1}&	{3.68}&{0.02}&	{M5.5}&	{1}&	{13}&{09}&	{6.61}&{0.02}&	{6.37}&{0.05}&	{6.01}&{0.02}&	{5.87}&{0.03$^{f}$}\\{GJ 54.1}&	{1}&	{3.72}&{0.04}&	{M4.5}&	{1}&	{12}&{10}&	{6.42}&{0.02}&	{6.17}&{0.04}&	{5.89}&{0.01}&	{5.72}&{0.04$^{f}$}\\{GJ 273}&	{1}&	{3.76}&{0.01}&	{M3.5}&	{1}&	{9}&{85}&	{4.86}&{0.02}&	{4.72}&{0.07}&	{4.46}&{0.02}&	{4.33}&{0.03$^{f}$}\\{SCR 1845-6357 A}&	{1}&	{3.85}&{0.02}&	{M8.5}&	{2}&	{17}&{40}&	{8.51}&{0.02}&	{8.14}&{0.02}&	{7.38}&{0.02}&	{7.08}&{0.07}\\{SO 0253+1652}&	{1}&	{3.85}&{0.01}&	{M7.0}&	{2}&	{15}&{14}&	{7.59}&{0.05}&	{7.32}&{0.03}&	{6.90}&{0.02}&	{6.72}&{0.08}\\{GJ 191}&	{1}&	{3.91}&{0.01}&	{M1.0}&	{4}&	{8}&{85}&	{5.05}&{0.02}&	{4.95}&{0.08}&	{4.72}&{0.01}&	{4.60}&{0.03$^{f}$}\\{DEN 1048-3956}&	{1}&	{4.02}&{0.02}&	{M8.5}&	{2}&	{17}&{39}&	{8.45}&{0.02}&	{8.10}&{0.02}&	{7.46}&{0.02}&	{7.23}&{0.09}\\{GJ 860 A/B}&	{2}&	{4.03}&{0.02}&	{M3.0 / M4.0}&	{1}&	{9}&{57}&	{4.78}&{0.03}&	{4.69}&{0.08}&	{4.30}&{0.01}&	{4.12}&{0.03}\\{GJ 234 A/B}&	{2}&	{4.09}&{0.02}&	{M4.5}&	{1}&	{11}&{12}&	{5.49}&{0.02}&	{5.29}&{0.07}&	{4.99}&{0.01}&	{4.84}&{0.03}\\{GJ 628}&	{1}&	{4.27}&{0.03}&	{M3.0}&	{1}&	{10}&{10}&	{5.08}&{0.02}&	{4.91}&{0.07}&	{4.67}&{0.01}&	{4.57}&{0.03$^{f}$}\\{GJ 1}&	{1}&	{4.34}&{0.02}&	{M1.5}&	{3}&	{8}&{54}&	{4.52}&{0.02}&	{4.36}&{0.09}&	{4.20}&{0.01}&	{4.07}&{0.02$^{f}$}\\{GJ 473 A/B}&	{2}&	{4.39}&{0.09}&	{M5.5}&	{1}&	{12}&{49}&	{6.04}&{0.02}&	{5.79}&{0.05}&	{5.49}&{0.01}&	{5.35}&{0.04}\\{GJ 83.1}&	{1}&	{4.45}&{0.06}&	{M4.5}&	{1}&	{12}&{31}&	{6.65}&{0.02}&	{6.44}&{0.04}&	{6.10}&{0.01}&	{5.96}&{0.04$^{f}$}\\{GJ 687}&	{1}&	{4.54}&{0.02}&	{M3.0}&	{1}&	{9}&{17}&	{4.55}&{0.02}&	{4.40}&{0.09}&	{4.18}&{0.02}&	{4.06}&{0.02$^{f}$}\\{LHS 292}&	{1}&	{4.54}&{0.07}&	{M7.0}&	{2}&	{15}&{73}&	{7.93}&{0.03}&	{7.71}&{0.02}&	{7.30}&{0.02}&	{7.03}&{0.09$^{f}$}\\{GJ 1245 A/C}&	{2}&	{4.54}&{0.02}&	{M5.5}&	{1}&	{13}&{41}&	{6.85}&{0.02}&	{6.60}&{0.07}&	{6.24}&{0.02}&	{6.08}&{0.05}\\{GJ 1245 B}&	{1}&	{4.54}&{0.02}&	{M6.0}&	{1}&	{14}&{01}&	{7.39}&{0.02}&	{7.18}&{0.07}&	{6.85}&{0.02}&	{6.77}&{0.09$^{f}$}\\{GJ 674}&	{1}&	{4.54}&{0.03}&	{M2.5}&	{3}&	{9}&{37}&	{4.86}&{0.02}&	{4.71}&{0.08}&	{4.50}&{0.02}&	{4.34}&{0.03$^{f}$}\\{GJ 876 A}&	{1}&	{4.66}&{0.01}&	{M3.5}&	{1}&	{10}&{18}&	{5.01}&{0.02}&	{4.84}&{0.08}&	{4.64}&{0.01}&	{4.54}&{0.03$^{f}$}\\{GJ 1002}&	{1}&	{4.69}&{0.08}&	{M5.5}&	{1}&	{13}&{77}&	{7.44}&{0.02}&	{7.18}&{0.03}&	{6.86}&{0.02}&	{6.77}&{0.08$^{f}$}\\{LHS 288}&	{1}&	{4.77}&{0.06}&	{M5.5}&	{2}&	{13}&{92}&	{7.73}&{0.03}&	{7.50}&{0.03}&	{7.09}&{0.02}&	{6.75}&{0.07$^{f}$}\\{GJ 412 A}&	{1}&	{4.86}&{0.02}&	{M1.0}&	{1}&	{8}&{77}&	{4.77}&{0.02}&	{4.64}&{0.09}&	{4.46}&{0.01}&	{4.36}&{0.02$^{f}$}\\{GJ 412 B}&	{1}&	{4.86}&{0.02}&	{M5.5}&	{1}&	{14}&{44}&	{7.84}&{0.03}&	{7.61}&{0.02}&	{7.22}&{0.02}&	{7.12}&{0.09$^{f}$}\\{GJ 388}&	{1}&	{4.89}&{0.07}&	{M4.0}&	{4}&	{9}&{29}&	{4.59}&{0.02}&	{4.42}&{0.09}&	{4.27}&{0.02}&	{4.15}&{0.03$^{f}$}\\{GJ 832}&	{1}&	{4.95}&{0.02}&	{M1.5}&	{3}&	{8}&{66}&	{4.50}&{0.02}&	{4.29}&{0.08}&	{4.16}&{0.01}&	{4.05}&{0.02$^{f}$}\\{LP 944-020}&	{1}&	{4.97}&{0.10}&	{M9.0}&	{2}&	{18}&{69}&	{9.55}&{0.02}&	{9.13}&{0.02}&	{8.27}&{0.02}&	{8.00}&{0.11$^{f}$}\\{GJ 166 C}&	{1}&	{4.98}&{0.01}&	{M4.5}&	{3}&	{11}&{24}&	{5.96}&{0.03}&	{5.81}&{0.05}&	{5.48}&{0.02}&	{5.34}&{0.04$^{f}$}\\{GJ 873}&	{1}&	{5.05}&{0.02}&	{M3.5}&	{1}&	{10}&{22}&	{5.30}&{0.02}&	{5.24}&{0.06}&	{4.89}&{0.02}&	{4.75}&{0.03}\\{GJ 682}&	{1}&	{5.05}&{0.05}&	{M3.5}&	{3}&	{10}&{95}&	{5.61}&{0.02}&	{5.35}&{0.07}&	{5.20}&{0.02}&	{5.13}&{0.04$^{f}$}\\{GJ 1116 A/B}&	{2}&	{5.23}&{0.07}&	{M5.5}&	{1}&	{13}&{65}&	{6.89}&{0.02}&	{6.64}&{0.04}&	{6.28}&{0.02}&	{6.21}&{0.08}\\{G 099-049}&	{1}&	{5.24}&{0.05}&	{M3.5}&	{1}&	{11}&{31}&	{6.04}&{0.02}&	{5.94}&{0.05}&	{5.61}&{0.02}&	{5.47}&{0.04}\\{LHS 1723}&	{1}&	{5.32}&{0.04}&	{M4.0}&	{3}&	{12}&{22}&	{6.74}&{0.02}&	{6.53}&{0.05}&	{6.23}&{0.02}&	{6.13}&{0.05}\\{GJ 445}&	{1}&	{5.34}&{0.05}&	{M3.5}&	{1}&	{10}&{79}&	{5.95}&{0.03}&	{5.75}&{0.06}&	{5.51}&{0.02}&	{5.38}&{0.03}\\{GJ 526}&	{1}&	{5.41}&{0.02}&	{M1.5}&	{1}&	{8}&{46}&	{4.42}&{0.02}&	{4.37}&{0.10}&	{4.19}&{0.01}&	{4.10}&{0.02}\\{GJ 169.1 A}&	{1}&	{5.54}&{0.02}&	{M4.5}&	{5}&	{11}&{04}&	{5.72}&{0.02}&	{5.54}&{0.06}&	{5.32}&{0.01}&	{5.20}&{0.03}\\{GJ 251}&	{1}&	{5.61}&{0.05}&	{M3.0}&	{1}&	{10}&{02}&	{5.28}&{0.02}&	{5.16}&{0.07}&	{4.92}&{0.02}&	{4.78}&{0.03}\\{2MA 1835+3259}&	{1}&	{5.67}&{0.02}&	{M8.5}&	{6}&	{18}&{27}&	{9.17}&{0.02}&	{8.80}&{0.02}&	{8.16}&{0.02}&	{7.89}&{0.13}\\{GJ 205}&	{1}&	{5.68}&{0.03}&	{M1.5}&	{1}&	{7}&{95}&	{4.04}&{0.26}&	{3.74}&{0.12}&	{3.70}&{0.02}&	{3.63}&{0.02}\\{LP 816-060}&	{1}&	{5.71}&{0.11}&	{M4.0}&	{7}&	{11}&{50}&	{6.20}&{0.02}&	{6.02}&{0.05}&	{5.77}&{0.02}&	{5.63}&{0.04}\\{GJ 229 A}&	{1}&	{5.75}&{0.03}&	{M1.0}&	{1}&	{8}&{14}&	{4.17}&{0.23}&	{4.06}&{0.13}&	{3.97}&{0.02}&	{3.82}&{0.04}\\{GJ 213}&	{1}&	{5.83}&{0.03}&	{M4.0}&	{1}&	{11}&{57}&	{6.39}&{0.02}&	{6.23}&{0.05}&	{5.90}&{0.02}&	{5.68}&{0.04}\\{GJ 693}&	{1}&	{5.84}&{0.08}&	{M2.0}&	{3}&	{10}&{76}&	{6.02}&{0.02}&	{5.92}&{0.06}&	{5.60}&{0.02}&	{5.48}&{0.04}\\{GJ 752 A}&	{1}&	{5.85}&{0.02}&	{M3.0}&	{1}&	{9}&{10}&	{4.67}&{0.02}&	{4.47}&{0.08}&	{4.38}&{0.01}&	{4.27}&{0.03}\\{GJ 752 B}&	{1}&	{5.85}&{0.02}&	{M8.0}&	{1}&	{17}&{45}&	{8.77}&{0.02}&	{8.47}&{0.02}&	{8.08}&{0.02}&	\multicolumn{2}{c}{N/A}\\{GJ 570 B/C}&	{2}&	{5.86}&{0.02}&	{M1.0}&	{1}&	{8}&{09}&	{3.80}&{0.23}&	{4.06}&{0.06}&	{3.74}&{0.02}&	{3.62}&{0.03}\\{GJ 754}&	{1}&	{5.91}&{0.05}&	{M4.5}&	{3}&	{12}&{23}&	{6.85}&{0.03}&	{6.65}&{0.03}&	{6.36}&{0.02}&	{6.20}&{0.06}\\{GJ 588}&	{1}&	{5.93}&{0.04}&	{M2.5}&	{3}&	{9}&{31}&	{4.76}&{0.02}&	{4.70}&{0.07}&	{4.46}&{0.01}&	{4.36}&{0.03}\\{GJ 1005 A/B}&	{2}&	{5.94}&{0.03}&	{M4.0}&	{1}&	{11}&{49}&	{6.39}&{0.02}&	{6.17}&{0.04}&	{5.91}&{0.02}&	{5.81}&{0.04}\\{GJ 908}&	{1}&	{5.95}&{0.04}&	{M1.0}&	{1}&	{8}&{98}&	{5.04}&{0.02}&	{5.02}&{0.07}&	{4.76}&{0.02}&	{4.66}&{0.03}\\{GJ 285}&	{1}&	{5.98}&{0.07}&	{M4.0}&	{1}&	{11}&{19}&	{5.70}&{0.02}&	{5.51}&{0.06}&	{5.29}&{0.01}&	{5.13}&{0.03}\\\noalign{\smallskip}
\hline
\end{tabular}
\end{table*}

\addtocounter{table}{-1}

\begin{table*}[ht!]
\caption{Observational data (contd.)}
\centering
\begin{tabular}{lcr@{$\pm$}lccr@{.}lr@{$\pm$}lr@{$\pm$}lr@{$\pm$}lr@{$\pm$}lr@{$\pm$}lr@{$\pm$}l} 
\hline\hline\noalign{\smallskip}
{}&        {stars in}&        \multicolumn{2}{c}{Distance}&	{Spectral}&	{SpTy}&    \multicolumn{2}{c}{V}&        \multicolumn{2}{c}{K$_s$}&        \multicolumn{2}{c}{$F_{3.4}$}&        \multicolumn{2}{c}{$F_{12}$}&        \multicolumn{2}{c}{$F_{22}$} \\
{Name} &  {beam} & \multicolumn{2}{c}{(pc)$^{a}$}& 	{Type$^{b}$}&	{ref.}&           \multicolumn{2}{c}{(mag)$^{c}$}&    \multicolumn{2}{c}{(mag)$^{d}$}&    \multicolumn{2}{c}{(mag)$^{e}$}&    \multicolumn{2}{c}{(mag)$^{e}$}&    \multicolumn{2}{c}{(mag)$^{e}$}\\
\noalign{\smallskip}
\hline
\noalign{\smallskip}
{GJ 268 A/B}&	{2}&	{6.12}&{0.07}&	{M4.5}&	{1}&	{11}&{48}&	{5.85}&{0.02}&	{5.69}&{0.06}&	{5.38}&{0.02}&	{5.20}&{0.03}\\
{GJ 784}&	{1}&	{6.20}&{0.04}&	{K7.0}&	{3}&	{7}&{95}&	{4.28}&{0.02}&	{4.15}&{0.10}&	{4.05}&{0.02}&	{3.98}&{0.02}\\{GJ 555}&	{1}&	{6.22}&{0.08}&	{M3.5}&	{1}&	{11}&{32}&	{5.94}&{0.03}&	{5.79}&{0.06}&	{5.54}&{0.02}&	{5.41}&{0.03}\\{GJ 896 A/B}&	{2}&	{6.25}&{0.06}&	{M3.5 / M4.5}&	{1}&	{10}&{10}&	{4.94}&{0.03}&	{4.76}&{0.08}&	{4.64}&{0.02}&	{4.59}&{0.03}\\{GJ 581}&	{1}&	{6.34}&{0.07}&	{M2.5}&	{1}&	{10}&{57}&	{5.84}&{0.02}&	{5.69}&{0.06}&	{5.48}&{0.01}&	{5.33}&{0.03}\\{LHS 2090}&	{1}&	{6.37}&{0.11}&	{M6.0}&	{2}&	{16}&{10}&	{8.44}&{0.02}&	{8.23}&{0.03}&	{7.80}&{0.02}&	{7.96}&{0.22}\\{LHS 337}&	{1}&	{6.38}&{0.08}&	{M4.5}&	{3}&	{12}&{75}&	{7.39}&{0.02}&	{7.24}&{0.03}&	{6.86}&{0.02}&	{6.72}&{0.05}\\{GJ 661 A/B}&	{2}&	{6.40}&{0.05}&	{M3.0}&	{1}&	{9}&{37}&	{4.83}&{0.02}&	{4.71}&{0.08}&	{4.46}&{0.02}&	{4.34}&{0.02}\\{G 180-060}&	{1}&	{6.41}&{0.16}&	{M5.0}&	{3}&	{14}&{76}&	{8.51}&{0.02}&	{8.29}&{0.02}&	{7.95}&{0.02}&	{7.90}&{0.13}\\{GJ 644 A/B/D}&	{3}&	{6.45}&{0.02}&	{M2.5}&	{1}&	{9}&{03}&	{4.40}&{0.04}&	{4.22}&{0.09}&	{4.04}&{0.02}&	{3.90}&{0.03}\\{GJ 644 C}&	{1}&	{6.45}&{0.02}&	{M7.0}&	{1}&	{16}&{78}&	{8.82}&{0.02}&	{8.59}&{0.02}&	{8.13}&{0.02}&	{7.86}&{0.18$^{f}$}\\{GJ 643}&	{1}&	{6.45}&{0.02}&	{M3.5}&	{1}&	{11}&{80}&	{6.72}&{0.02}&	{6.57}&{0.03}&	{6.29}&{0.02}&	{6.12}&{0.05}\\{GJ 625}&	{1}&	{6.53}&{0.04}&	{M1.5}&	{1}&	{10}&{10}&	{5.83}&{0.02}&	{5.68}&{0.06}&	{5.46}&{0.02}&	{5.33}&{0.02}\\{GJ 1128}&	{1}&	{6.53}&{0.10}&	{M4.5}&	{1}&	{12}&{74}&	{7.04}&{0.02}&	{6.80}&{0.03}&	{6.51}&{0.01}&	{6.40}&{0.03}\\{GJ 1156}&	{1}&	{6.54}&{0.13}&	{M5.0}&	{1}&	{13}&{80}&	{7.57}&{0.03}&	{7.33}&{0.03}&	{6.99}&{0.02}&	{6.91}&{0.08}\\{LHS 3003}&	{1}&	{6.56}&{0.09}&	{M7.0}&	{1}&	{17}&{05}&	{8.93}&{0.03}&	{8.69}&{0.02}&	{8.27}&{0.02}&	{8.12}&{0.27$^{f}$}\\{GJ 408}&	{1}&	{6.70}&{0.07}&	{M2.5}&	{3}&	{10}&{02}&	{5.50}&{10.00}&	{5.38}&{0.06}&	{5.19}&{0.01}&	{5.09}&{0.03}\\{GJ 829 A/B}&	{2}&	{6.71}&{0.08}&	{M3.5}&	{1}&	{10}&{30}&	{5.45}&{0.02}&	{5.30}&{0.07}&	{5.06}&{0.01}&	{4.90}&{0.03}\\{G 041-014 A/B/C}&	{3}&	{6.77}&{0.09}&	{M3.5}&	{3}&	{11}&{46}&	{5.69}&{0.02}&	{5.49}&{0.06}&	{5.25}&{0.02}&	{5.08}&{0.03}\\{GJ 402}&	{1}&	{6.80}&{0.14}&	{M4.0}&	{1}&	{11}&{65}&	{6.37}&{0.02}&	{6.26}&{0.05}&	{5.98}&{0.02}&	{5.86}&{0.04}\\{GJ 880}&	{1}&	{6.83}&{0.04}&	{M1.5}&	{1}&	{8}&{65}&	{4.52}&{0.02}&	{4.43}&{0.08}&	{4.33}&{0.01}&	{4.22}&{0.02}\\{GJ 299}&	{1}&	{6.84}&{0.14}&	{M4.0}&	{1}&	{12}&{82}&	{7.66}&{0.03}&	{7.44}&{0.03}&	{7.12}&{0.02}&	{7.01}&{0.09}\\

\noalign{\smallskip}
\hline
\end{tabular}
\tablefoot{\\
\tablefoottext{a}{Distances calculated from RECONS parallaxes \citep{Henry2006}.}\\
\tablefoottext{b}{Spectral types from the literature. When only one spectral type is given for a binary / multiple system, this represents the spectral type of the whole system and the individual spectral types are unknown. The sources of the spectral type are given in the next column and are: 1) \citealp{Henry2002}, 2) \citealp{Henry2004}, 3) \citealp{Hawley1996}, 4) \citealp{Keenan1989}, 5) \citealp{Gray2003}, 6) \citealp{Reid2003}, 7) \citealp{Gray2006}.}\\
\tablefoottext{c}{$V$ magnitudes taken from the RECONS database; where more than one star is in the \emph{WISE} / 2MASS beam, the fluxes are added.}\\
\tablefoottext{d}{$K_s$ magnitudes and errors are taken from the 2MASS database \citep{Skrutskie2006}; where more than one star is in the \emph{WISE} beam, the fluxes of all stars in the beam are added.}\\
\tablefoottext{e}{\emph{WISE} bands w1 centered at 3.4$\mu$m, w3 centered at 12$\mu$m, and w4 centered at 22$\mu$m \citep{Wright2010}.}\\
\tablefoottext{f}{Stars with 24 $\mu$m \emph{Spitzer} flux measurements from \citet{Gautier2007}.}
}
\end{table*}

Table \ref{table:observationaldata} shows the observational data for these 85 sources. The \emph{WISE} data is complemented with $K_s$ magnitudes from the 2MASS survey and spectral types from the literature as well as $V$ band data and distance information from RECONS. If multiple stars were present in the \emph{WISE} beam, the V magnitudes were calculated by combining the single-source V band magnitudes from RECONS. In the single case that stars were unresolved in \emph{WISE}, but resolved with 2MASS (GJ 896 A/B), the 2MASS magnitudes were calculated in the same way. In all other cases, the sources were confused in 2MASS as well.

For the purposes of this paper, the data from three of the four \emph{WISE} bands were used: The first, third, and fourth. The second band, centered at 4.6 $\mu$m, was omitted. The reason is that while the w4 band is just about able to pick up the faintest and farthest dwarfs in our sample at magnitudes beyond 8, the \emph{WISE} measurements for the earlier bands run into saturation at magnitudes brighter than $\sim$6. While the first \emph{WISE} band behaves quite well and high-quality magnitudes can be recovered, the magnitudes of the second band are prone to significant errors for bright stars. Because of this, we did not trust the 4.6$\mu$m measurements for magnitudes smaller than $\sim$6 (in our case, absolute magnitudes smaller than $\sim$8-9), which means large parts of our stellar sample. We illustrate this in Figure \ref{fig:SecondChannelBad}. Because of this, we focused on the 3.4, 12, and 22 $\mu$m bands. 

\subsubsection{WISE data quality}
\label{sec:WISEquality}

\begin{figure}
\includegraphics[width=9cm]{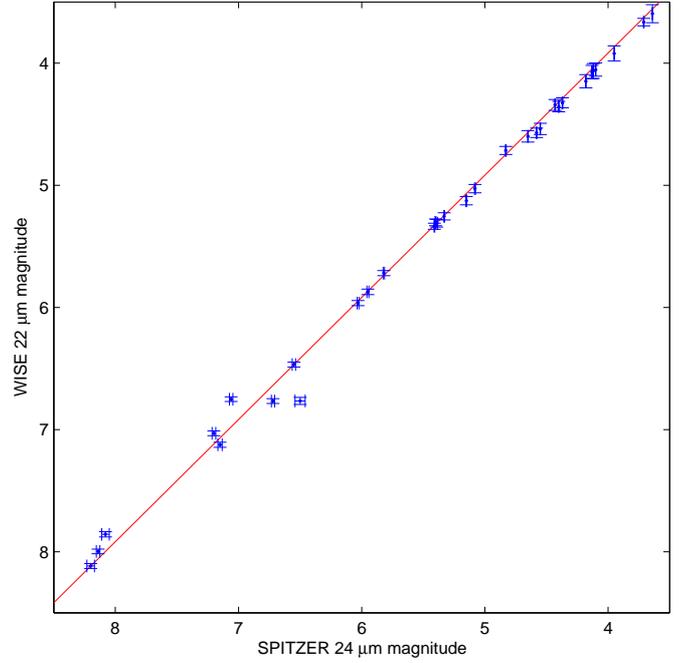}
\caption{Comparison of \emph{WISE} and \emph{Spitzer} data for the sources in common with \citet{Gautier2007}. The red line shows a weighted least-squres fit for the offset between the \emph{Spitzer} and \emph{WISE} magnitudes, which is 0.083 mag with the WISE magnitudes being fainter. This offset stems from the different wavelength passbands (centered at 22 $\mu$m vs. 24 $\mu$m). The \emph{WISE} data match the \emph{Spitzer} data very well with only a few outliers at the faint end.}
\label{fig:SpitzervsWISE}
\end{figure}

The \emph{WISE} survey gives, in addition to the magnitudes in the four bands, error estimates for these four magnitudes. To check the data as well as the error estimates, we compared the w4 (22 $\mu$m) band data with data from \citet{Gautier2007}, who conducted a similar study using observations with the \emph{Spitzer} space telescope. Some of our samples overlap (the corresponding stars are marked with ($^{e}$) in Table \ref{table:observationaldata}). \emph{Spitzer}, providing pointed observations, performs significantly better than \emph{WISE} in measuring infrared magnitudes. For the 34 stars of which both \emph{Spitzer} 24$\mu$m and \emph{WISE} 22$\mu$m data ARE available, the \emph{Spitzer} formal uncertainties (not including systematic calibration uncertainties, c.f. \citealp{Gautier2007}) are consistently smaller, often by a factor exceeding 10. We can therefore use the \emph{Spitzer} data as a benchmark for the \emph{WISE} data, as we did in Figure \ref{fig:SpitzervsWISE}. There is a slight offset between the \emph{Spitzer} and \emph{WISE} magnitudes that we determined to be 0.083 magnitudes using a weighted least-squares fit. For the high-luminosity end, the fit is very good and suggests that the \emph{WISE} errors might actually be overestimated (or there is a fraction in these errors that represents the global calibration uncertainty, which has no influence on this comparison other than a global offset, which we corrected for). At the low-luminosity end, there are some outliers. At least one of these outliers can be explained because visual inspection shows that the \emph{WISE} source is not exactly centered on the actual source. We conclude from this that at least in the longest waveband, the data quality is mostly good, but some outliers might be expected. No such statement can be made about the other wave bands. However, as our analysis will show, the data quality is indeed excellent and the \emph{WISE} errors seem to be generally overestimated.

\section {Analysis}

For our 85 systems, we obtained magnitudes in seven different bands ($V$, the \emph{2MASS} bands $J$, $H$ and $K_s$ as well as the \emph{WISE} bands w1, w3 and w4). In addition, we also have parallax and thus distance information, giving us access to the absolute magnitudes. The difficulty we face with M-dwarfs is that, unlike solar-type or A-stars, they are not well into their Rayleigh-Jeans tail at 3.4 $\mu$m. This means that not only the $J$-$H$, $J$-$K_s$ and $H$-$K_s$ colors depend on spectral type, but that the same applies for the colors in the \emph{WISE} bands.

\subsection {A proxy for stellar temperature}

\begin{figure*}
\includegraphics[width=18cm]{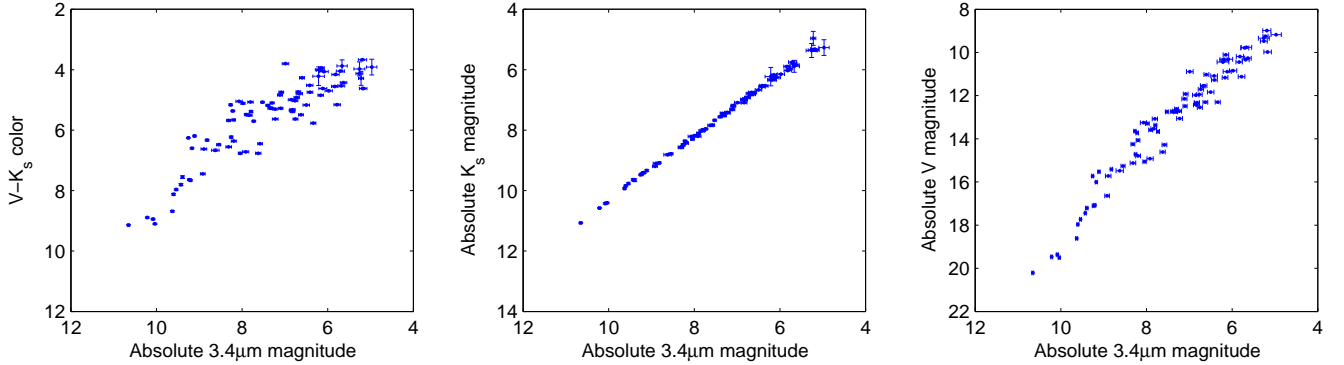}
\caption{Comparison of WISE absolute magnitudes in w1 (3.4 $\mu$m) band with $V$-$K_s$ color (left panel). As can be seen, the two proxies for stellar temperature do not match well. This is not because of the $K_s$ magnitude (as can be seen in the middle panel), but because of the magnitude in the visual band (right panel). 
An empirical determination of M-dwarf mid-infrared colors as a function of the $V$-$K_s$ color index can be found in the appendix.}
\label{fig:w1vsVKs}
\end{figure*}

Because we expect the dependence of the \emph{WISE} colors for stars without excess to be mainly with the effective stellar temperature $T_{eff}$, we need a good proxy for this. \citet{Casagrande2008} have shown that none of the \emph{2MASS} colors ($J$-$H$, $J$-$K_s$, $H$-$K_s$) are good proxies for effective stellar temperature, but show significant scatter and are not necessarily a monotonic function of spectral type. Good proxies are the $V$-$K_s$ color as well as absolute magnitudes in a variety of bands.

The spectra of M-dwarfs are dominated in the visible TiO bands. The $V$-band flux for a given $T_{eff}$ may therefore significantly depend on abundances and atmospheric dynamics. Furthermore, many M-dwarfs are known to be variable in the visible band (e.g., \citealp{Hawley1993}), which in turn can influence the $V$-$K_s$ color, depending on at which point in the variability cycle the $V$-band data was taken. To assess this, we plot the absolute 3.4 $\mu$m magnitude obtained from \emph{WISE} against the $V$-$K_s$ color in Figure \ref{fig:w1vsVKs}. If both are indeed good proxies for stellar temperature, we expect to be able to fit a (not necessarily straight) line through the data points. This is clearly not the case. Plots against the $K_s$ and $V$ absolute magnitude show that the main contribution to the scatter comes from the $V$ magnitude.

We conclude that because of strong molecular absorptions and variability in the $V$ band, the $V$-$K_s$ color is not an ideal proxy for the stellar temperature of late-type dwarfs. We chose to use the 3.4 $\mu$m absolute magnitude instead, because it is a good proxy for stellar temperature and furthermore can give us the best data consistency because it originates from the same instrument. The reduced $\chi^2$ goodness-of-fit measurements of our analysis support this choice because the fits to the data are better. 

There is one caveat to using absolute magnitudes as a proxy, though. Because pre-main-sequence stars are too bright for their spectral type due to their larger diameters, the predictions for their spectral types / surface temperatures and consequently their colors in the mid-infrared will be off (c.f. Section \ref{sec:AUMic}). For this paper, this is not a problem because there are no known pre-main-sequence stars in our sample. For young stars, however, it is better to use the $V$-$K_s$ color as a proxy, which 
is indeed also a very decent proxy for stellar temperature even for these late-type dwarfs. We show the fit against the $V$-$K_s$ color index for comparison and reference in the appendix.

\begin{figure*}[ht!]
\centering
\includegraphics[width=17.2cm]{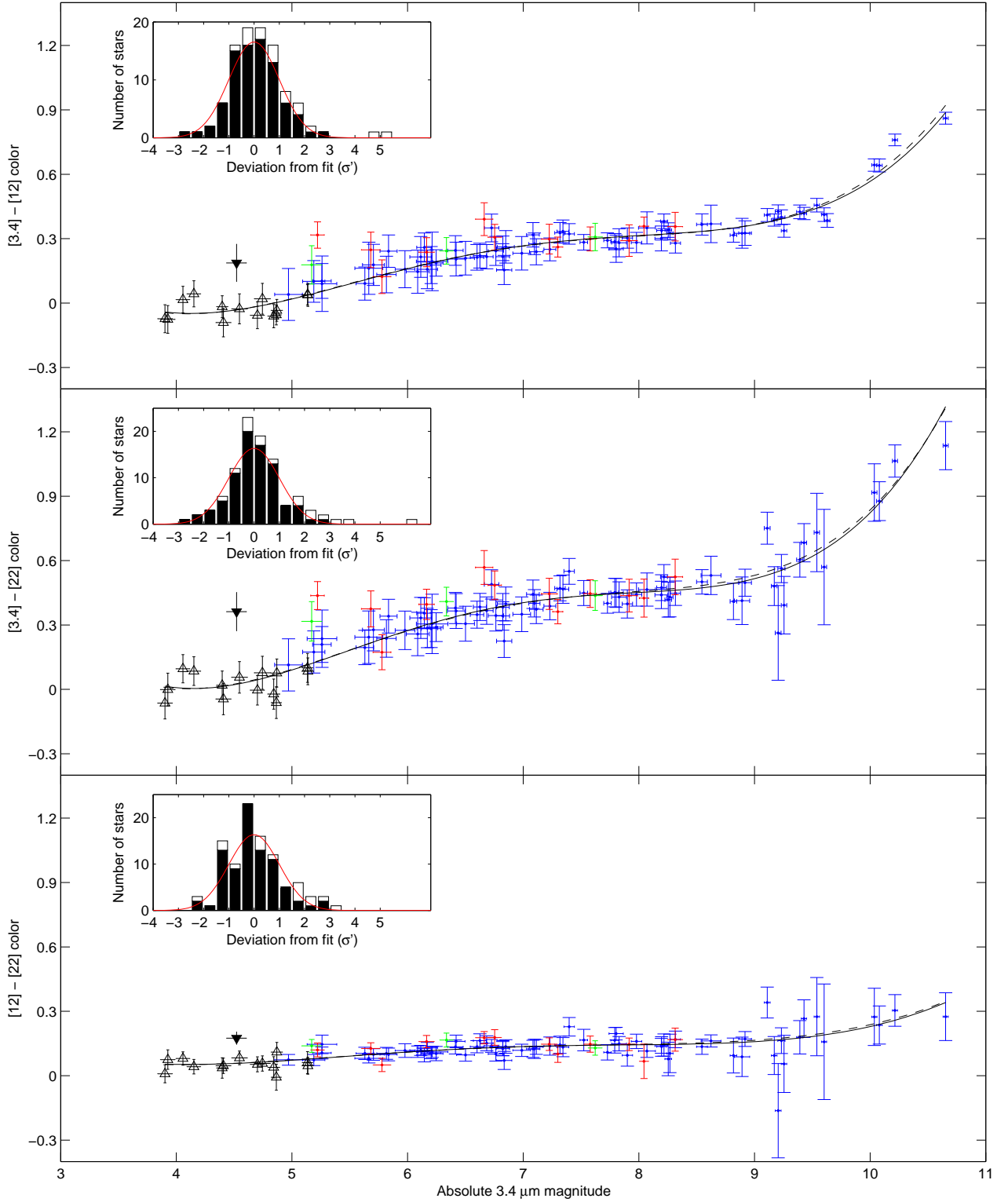}
\caption{WISE colors plotted against 3.4 $\mu$m absolute magnitude. Single M-dwarfs are blue, binary systems are marked in red, triple systems in green. The black open triangles show the K dwarfs included to better constrain the high-mass end of the fit. The black dashed line shows the third-order polynomial fit using all single dwarfs, while the black solid line shows the fit after iteratively removing outliers and is used for our analysis. The inset shows the histogram for the $\sigma'$ error value adopted for our analysis (see definition in the text). The solid bars represent the stars that were used in the fit, the blank stacked bars are for the stars that were omitted from the fit (multiples and outliers). In all three cases, the histograms are consistent with a standard-normal distribution (plotted here as a red line). For reference and as test of our technique, the black filled downward triangle to the left represents the well-known excess source AU Mic, which seems to have significant excess in all three colors. For a discussion of AU Mic, see Section \ref{sec:AUMic}.}
\label{fig:plots1}
\end{figure*}

\subsection {Variability in the WISE bands}

As we have discussed above, many M-dwarfs are known to be variable in the visible. Several of our stars are flare stars, producing short-term variability in the ultraviolet. 
Out of the 34 stars in our sample that were also observed with \emph{Spitzer} by \citet{Gautier2007}, the SIMBAD database marks more than half as flare and/or variable stars. While we cannot comment on the accuracy of these classifications, we see that flare and variable stars are part of this sample. On the other hand, we see that the \emph{WISE} and \emph{Spitzer} magnitudes match well (cf. Figure \ref{fig:SpitzervsWISE}) at the bright end, while outliers are present at the faint end. \citet{Tofflemire2012} observed three M-dwarfs for flares and short-term variability and found that while the flares are intense in the ultraviolet, the variability they produce in the near-infrared 2MASS regime is below 7-11 mmag depending on the band. From this we conclude that the outliers in the longer \emph{WISE} bands are mostly not due to variability, but have other reasons (for example, bad centering, background contaminations or contaminations from nearby sources).

We can get rid of most of the variability in the mid-infrared by using the \emph{WISE} colors. All four \emph{WISE} bands are taken at the same time, which means that any variability that influences all mid-infrared bands equally will not influence the \emph{WISE} colors at all. The \emph{WISE} catalog provides a variability flag for each of the four bands. Out of the 85 sources in our sample, only a small minority of seven are flagged as variable in at least one of the bands. We conclude that variability is likely to not be an issue for excess detection in our analysis for most of our sample.

\subsection {Multiplicity}
\label{sect:mult}

As we have noted before, \emph{WISE} is not able to resolve all known multiple systems within our sample. Because we do not want to exclude these systems from our study, we have to investigate what happens to the \emph{WISE} magnitudes and colors in multiple systems.

Equal-mass binaries will simply have double the luminosity of the single star, the colors should not be influenced. They move to the left by 0.75 magnitudes in Figure \ref{fig:plots1}. Binaries of different mass will move less in absolute magnitude (compared to their primary alone), but might move slightly upwards if the secondary contributes significantly and has a significantly different color. This effect gets smaller for a larger difference in spectral class and thus a larger color difference because the secondary will then only have a fraction of the brightness of the primary, and we expect these binaries to lie relatively close to the position their primary would occupy in the diagram. The same is in principle true for triple systems, even though somewhat larger deviations are to be expected.

Because of the above discussion, we do not use the multiple systems within our sample to obtain the polynomial fit. We do compare them to the fit, though - bearing in mind that we will have to treat them specially if any of them turns out to show signs for excess. Binary and triple systems are marked in Figure \ref{fig:plots1}. As we expect, the multiple systems lie preferentially above the fitted polynomial line.

\subsection{Fitting and intrinsic spread of mid-infrared colors}

In order to be able to detect excesses, we need to know the expected mid-infrared color for every source. Instead of using model stellar atmospheres, we choose a purely data-driven approach and use the data at hand (which might include excess sources) to fit for the expected mid-infrared colors. This is only possible if the number of data points significantly exceeds the number of parameters to fit. It also assumes that the number of excess sources is small compared to the general sample size, which is a reasonable assumption (c.f. section \ref{sec:comp}). We use a polynomial of 4th order to describe the mid-infrared colors and obtain the fit through an iterative fitting routine that removes outliers in each step (details on this fitting routine and its parameters can be found in the appendix). Outliers are defined as sources that diverge from the fit by more than 3$\sigma'$. $\sigma'$ is defined from the data and the fit as $$\sigma_i'=\frac{e_{i}}{\sigma_i} \frac{1}{\sqrt{\chi^2}},$$ where $\sigma_i$ is the \emph{WISE} error estimate for source i, $e_i$ is the deviation from the fit and $\chi^2$ is the overall goodness-of-fit measure of the fit. We note at this point that the iterative fitting procedure does not change the number of excess detections (i.e. there is no excess source that would not have been detected as excess source had we not used an iterative fit).

\citet{Gautier2007} stated that the limits for detecting mid-infrared excesses stem from the scatter in photospheric flux predictions from shorter wavelengths and that the mid-infrared properties of M-dwarfs are not well known. While we did not predict the mid-infrared fluxes, but instead directly fit the colors, we are also affected by a possible intrinsic spread in the mid-infrared colors, the amount of which is unknown. 
However, we know that if the intrinsic spread was negligible, our model was a reasonably good fit to the data (which visually seems the case) and the \emph{WISE} error estimates were exactly right, then the deviations from the fit divided by their respective \emph{WISE} error estimates ($\frac{e_{i}}{\sigma_i}$) should follow a standard-normal distribution and the $\chi^2$ measure should be $\sim1$. In fact, the $\chi^2$ measures we obtained were significantly smaller than one, which means that the intrinsic spread must be significantly smaller than the \emph{WISE} error estimates, and furthermore, the \emph{WISE} errors are presumably overestimated.

Because separating the errors from the (overestimated) \emph{WISE} measurements and the intrinsic spread would only be possible with a large set of (not necessarily justified) assumptions, we used the above-defined $\sigma'$ value instead. If the amount of overestimation is about the same for all \emph{WISE} error estimates and if the intrinsic spread is in fact much smaller than these estimates, $\sigma'$ follows a standard-normal distribution. We present histograms of $\sigma'$ in Figure \ref{fig:plots1} as insets. We also used the Kolmogorov-Smirnov test and the Shapiro-Wilk test to check whether the distributions are compatible with a Gaussian distribution. The results are shown in Table \ref{table:statisticalTests}. For all three examined colors, there are no hints that the distribution is not Gaussian.

It is worth noting that we fit the three colors individually and not simultaneously, which means that they are not self-consistent and the net color ([3]-[12] + [12]-[22] - [3]-[22]) is not zero. The residuals, however, are so low (0.005 magnitudes on average) that this has no influence on our excess detection technique.

\begin{table}
\caption{Statistical tests for $\sigma'$}
\label{table:statisticalTests}    
\centering
\begin{tabular}{ccc}
\hline\hline\noalign{\smallskip}
{Color index} &     {KS test}&{SW test}\\
{} &     {(p value)}&{(p value)}\\
\noalign{\smallskip}
\hline
\noalign{\smallskip}

{$[3.4]$-$[12]$}&{0.84}&{0.82}\\
{$[3.4]$-$[22]$}&{0.74}&{0.71}\\
{$[12]$-$[22]$}&{0.47}&{0.41}\\

\noalign{\smallskip}
\hline
\end{tabular}
\tablefoot{Probability of the data ($\sigma'$) being drawn from a Gaussian distribution as calculated using the KS (Kolmogorov-Smirnov) and SW (Shapiro-Wilk) tests. All data are consistent with being standard-normal distributed.}
\end{table}

\subsection{Excess detection}

Because of the Gaussian nature of $\sigma'$, it can directly be used as a measure to detect excess. We considered a source to show statistically significant excess if $\sigma_i'>3$. No source in any of the three colors shows a statistically significant deficit, defined as $\sigma_i'<-3$. This gave us confidence that significant excess would not be just a statistical outlier.

\subsection{Excess sources}

In our sample, we found two sources with $\sigma'$ values $>$ 3 in $[3.4]$-$[12]$ color, three in $[3.4]$-$[22]$ color and one in $[12]$-$[22]$ color. We list these sources in Table \ref{table:excesssources}. The six detections are for four individual stars. We discuss each star in detail.

\begin{table*}
\caption{Sources showing excess in our analysis - discussion in text}
\label{table:excesssources}    
\centering
\begin{tabular}{ccccl}
\hline\hline\noalign{\smallskip}
{Name} &     {$[3.4]$-$[12]$}&     {$[3.4]$-$[22]$}&     {$[12]$-$[22]$}&	{Comment}\\
\noalign{\smallskip}
\hline
\noalign{\smallskip}

{GJ 570 B/C}&    {\textbf{5.27}}&    {\textbf{6.05}}&    {1.73}&	{Contamination from nearby K4 primary}\\
{SCR 1845-6357 A}&    {\textbf{4.84}}&    {1.92}&    {0.87}&	{T6 brown dwarf companion in beam}\\
{LHS 288}&    {1.34}&    {\textbf{3.78}}&    {\textbf{3.49}}&		{22$\mu$m measurement strongly inconsistent with Spitzer measurement}\\
{GJ 860 A/B}&    {2.41}&    {\textbf{3.21}}&    {2.35}&			{Apparent excess caused by binarity}\\

\noalign{\smallskip}
\hline
\end{tabular}
\tablefoot{The four sources identified as excess sources from our $\sigma'>3$ test with their corresponding $\sigma'$ values for the three colors we constructed. Statistically significant excesses are marked in bold. The comment gives possible reasons why the detected excess could be spurious or not originate from a debris disk. The sources are discussed individually in the text.}
\end{table*}

\subsubsection{GJ 570 B/C}

The binary GJ 570 B/C shows excess at 12 and 22$\mu$m and is flagged as variable in all three considered bands. Visual inspection of the \emph{WISE} images shows a second nearby source. The GJ 570 system actually consists of four individual stars: The K4 primary GJ 570 A, the two close M-dwarfs GJ 570 B and C, and the T7 brown dwarf GJ 570 D. The brown dwarf is at large separation ($\sim250\arcsec$), the source next to the B/C binary is the A primary at a separation of about $20\arcsec$. Both sources are bright (3.8 magnitudes in $K_s$), and since they are close, contamination could be a problem. Indeed, a close examination of the $J$, $H$ and $K_s$ band in combination with the \emph{WISE} bands yields that the GJ 570 B/C is measured to be more than 0.25 magnitudes fainter in the first \emph{WISE} band than in $K_s$ and more than 0.3 magnitudes fainter in the first \emph{WISE} band compared to the third. We conclude that the most likely explanation is that the measurement in the first \emph{WISE} band is incorrect. This explains both the apparent excess in the third and in the fourth band, since the colors are calculated against the first \emph{WISE} band. It would not explain any excess in $[12]$-$[22]$, and indeed no excess is seen at this color (the excess value of $1.7\sigma'$ is well below our threshold of $3\sigma'$ and might be artificially enhanced because GJ 570 B/C is a binary, c.f. section \ref{sect:mult}). We therefore conclude that there is no significant sign for excess at either 12 or 22$\mu$m.

\subsubsection{SCR 1845-6357 A}

This single star shows a clear $4.8\sigma'$ excess at 12$\mu$m and no excess at 22$\mu$m, although the associated $\sigma'$ value is 1.9. The derived $\sigma'$ of 0.87 in $[12]$-$[22]$ is not significant. It is not flagged as variable in any of the WISE bands. The excess in magnitudes over the fitted line is 0.132, 0.144 and 0.064 magnitudes, corresponding to 12.9$\%$, 14.2$\%$ and 6.1$\%$ excess, respectively, with the caution that the latter two are not significant. 

The excess can be explained by the fact that SCR 1845-6357 has a T6 secondary at a separation of $\sim1\arcsec$ \citep{Biller2006, Kasper2007}. While T dwarfs usually do not contaminate the flux of a system sufficiently to produce a detectable excess in solar-type stars because of the large difference in emitted flux, SCR 1845-6357 A is a very late M-dwarf of spectral class M8.5 \citep{Henry2006} and thus itself very faint. The T dwarf is fainter by only $\sim$4 magnitudes in the H band \citep{Kasper2007}. While the flux in the 3.4$\mu$m band of \emph{WISE} is only on the level of $\sim1-2\%$ of the M-dwarf primary, it can explain the increased flux in the 12$\mu$m and 22$\mu$m bands. We conclude that the excess is real in both the 12 and 22$\mu$m bands, but does likely not originate from a debris disk, but from the T dwarf secondary. This shows the validity of our method, which is indeed able to pick up excesses in the 12$\mu$m band of as low as 13$\%$ for this very late dwarf.

\subsubsection{LHS 288}

LHS 288 is a single star showing excess in $[3.4]$-$[22]$ (3.8 $\sigma'$) and $[12]$-$[22]$ (3.5 $\sigma'$). There is no clear indication that the 22$\mu$m measurement of 6.75 magnitudes would be wrong, even though it can be seen in the \emph{WISE} image that there is a diffuse nebulosity in the background. The star is furthermore flagged as variable in the 3.4 and 22 $\mu$m \emph{WISE} bands. It is worth noting, however, that this source was also measured with \emph{Spitzer} by \citet{Gautier2007}. The two measurements are strongly inconsistent, with the \emph{WISE} measurement being 9.5 $\sigma$ above the \emph{Spitzer} measurement at 24$\mu$m. We conclude that the most likely explanation for LHS 288 is a poor measurement in the last \emph{WISE} band and not a real excess.

\subsubsection{GJ 860 A/B}
\label{sec:GJ860}

The close M3/M4 binary GJ 860 shows a derived excess at the longest \emph{WISE} band of $3.2\sigma'$. This corresponds to 0.254 magnitudes or 26.3$\%$ excess at this wavelength. At 12$\mu$m, the excess is at a $2.6\sigma'$ level (corresponding to 0.184 magnitudes / 18.4$\%$), the signal in $[12]$-$[22]$ color is $2.3\sigma'$ corresponding to 0.067 magnitudes or 6.4$\%$. The \emph{WISE} catalog flags this source as variable, but only in the third (12$\mu$m) band. The system does lie in a crowded region with possible background cirrus, but it is bright itself (magnitude 4.7 at 3.4$\mu$m, magnitude 4.1 at 22$\mu$m). Other bright sources are nearby, especially one source that is strong at 22$\mu$m but weak at the other wavelengths (in comparison to GJ 860). This source lies approximately 90$\arcsec$ to the southwest. It is unlikely that it strongly affects the \emph{WISE} measurement at this wavelength. Because this is a binary star, special precautions are necessary. The increased luminosity of the total system will move the star to the left in our diagram, and because the later-type member of the system will show a more red $[3.4]$-$[22]$ color, it will shift slightly upward compared to a single M3 dwarf. Assuming that the M3 dwarf is stronger by a factor of 2 in total luminosity, both effects can be calculated and the position of the M3 dwarf alone in the diagram can be derived. Doing so, we derive new values for $\sigma'$ of 2.3, 1.7 and 1.9, respectively. None of these are significant, and we therefore did not include this source in our excess sources. An excess might be present at levels below our detection threshold.

\subsection{Excess detection limits}

Because we handled each star individually rather than setting a general limit on $F_{IR}/F_{\star}$, with $F_{IR}$ being the detected infrared flux and $F_{\star}$ the expected stellar flux from our model fit, we did not obtain a global value of an excess detection limit for all our stars. Instead, we were able to detect ($3\sigma'$ limit) infrared flux excesses down to 10$\%$ in some stars while in other stars, even a 30$\%$ excess would not be detected due to a larger \emph{WISE} measurement error. Because of this, we calculated the excess detection limit for each star individually for each color assessed from its \emph{WISE} error and the $\chi^2$ value of the corresponding color fit. The typical (median) limits we derived are 14.9$\%$ for the $[3.4]$-$[12]$ excess, 17.9$\%$ for the $[3.4]$-$[22]$ excess and 7.0$\%$ for the $[12]$-$[22]$ excess. The mean values are higher due to a few sources for which we derived quite poor detection limits and measure 15.4$\%$, 20.2$\%$ and 10.9$\%$, respectively.

\begin{figure}
\includegraphics[width=9cm]{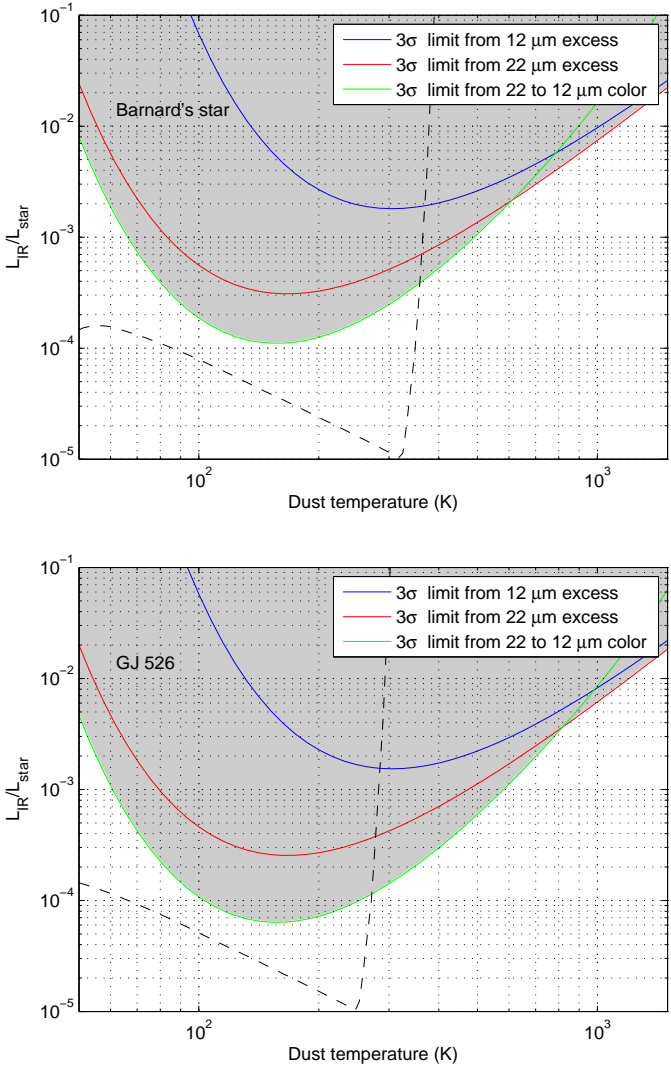}
\caption{Inferred dust limits for GJ 699 (Barnard's star, top) and GJ 526 (bottom). The different lines show the limits from the different \emph{WISE} colors. The gray shaded region is excluded by the combination of all three individual limits. For reference, the black dashed line shows the expected performance of SPHERE/ZIMPOL for a thin dust ring (see section \ref{sect:SPHERE}).}
\label{fig:ProxCen}
\end{figure}

These limits can be converted into limits on the fractional dust luminosity $L_{IR}/L_{\star}$ if we assume a single dust ring of a given temperature (see Fig. \ref{fig:ProxCen}). As can be seen, we put no strong constraints on the fractional dust luminosity for cold dust of temperatures lower than $\sim$50K. Large amounts of cold dust could be present around the M-dwarfs we observed to have no significant amount of excess at 12 and 22$\mu$m. At our most sensitive point, which is in the interesting (because of its link to terrestrial planet formation zones) 100-300K range, we derived 3$\sigma$ upper limits for $L_{IR}/L_{\star}$ of nearly $10^{-4}$ for Barnard's star (M3.5, 1.8pc) and nearly $6*10^{-5}$ for GJ 526 (M1, 5.4pc). These are indeed fairly typical values for our survey, although they depend slightly on the spectral type - for the earliest stars in our sample, we can go deeper by a factor of $\sim$2, while for the latest stars we are less deep, again by a factor of $\sim$2 (both compared to Barnard's star). This is because it becomes increasingly easy to detect dust when the difference in temperature between the dust and the host star is larger \citep{Backman1993}. Also because of this, following the discussion in \citet{Gautier2007}, we realized that while the limits we obtained for the fractional dust luminosity are less stringent than have been derived for earlier-type stars, this indeed corresponds to more stringent constraints on the dust mass at these temperatures (see discussion in section \ref{sect:masses}).

In most of the cases, the $[12]$-$[22]$ color is in the most sensitive to dust of these temperatures because such dust produces no significant excess at 12$\mu$m and because our excess limits are the most stringent ones in this color.

\subsection{AU Mic}
\label{sec:AUMic}
Although not in our sample due to its distance, it is interesting to investigate in passing the most well-known M-dwarf with a debris disk, AU Mic, an $\sim$12 Myr old M1e star in the $\beta$ Pictoris moving group \citep{Zuckerman2001}. While excess emission has been found at wavelengths of 70$\mu$m and beyond, no excess has been reported at mid- infrared wavelengths of 24$\mu$m or shorter \citep{Liu2004, Plavchan2009}. We plot it together with the stars from our sample in Figure \ref{fig:plots1}, where it can be seen to have significant excess in all colors (3.0$\sigma'$, 4.9$\sigma'$ and 5.8$\sigma'$, respectively). However, it lies in the K dwarf regime w.r.t. its absolute 3.4$\mu$m magnitude. This is because it has a larger surface area than the average M1 dwarf due to its young age. As we have seen, the M-dwarf mid-infrared colors depend on their spectral type / surface temperature, and the absolute 3.4$\mu$m magnitude is a poor proxy for either in case of pre-main-sequence stars that are too bright for their spectral type.

It is accordingly better to use the $V$-$K_s$ color as the proxy for stellar temperature for such young stars. Doing so, performing the fits and comparing the colors of AU Mic to them, we obtained $\sigma'$ values of 0.3, 1.2 and 3.3, respectively, for the $[3.4]$-$[12]$, $[3.4]$-$[22]$ and $[12]$-$[22]$ colors. This is significant in the last color, pointing to an excess of 6.7$\pm$2.0\% over the photosphere at 22$\mu$m. This can be compared to \citet{Plavchan2009}, who did not find a significant excess (3.5\% above the photosphere, 1.65$\sigma$ using \emph{Spitzer MIPS}) and \citet{Simon2012}, who did not detect an excess at 22$\mu$m either using the \emph{WISE preliminary release catalog} and a fixed threshold of $[12]$-$[22]$$>$1.0, stating that the AU Mic debris disk is too cold for detection with WISE.

This means that no mid-infrared excess is known for AU Mic, but both these results are consistent with its disk having an excess at 22/24$\mu$m, albeit a small one. The \emph{Spitzer} MIPS measurement of 155.2$\pm$3.2 mJy can be converted to 4.160$\pm$0.028 magnitudes using the zero points from the \emph{Spitzer MIPS Instrument Handbook}. This is consistent with the \emph{WISE} measurement of 4.137$\pm$0.025 magnitudes, both with and without taking into account the general offset between the \emph{Spitzer} and \emph{WISE} measurements discussed in Section \ref{sec:WISEquality}. This means that we only detect the small excess because our method is more sensitive. Indeed, comparing the two bands of \emph{WISE} directly using a calibration from other stars might currently be the most sensitive way to detect very small excesses.

Additional hints of an excess at shorter wavelengths than previously thought for AU Mic come from an unpublished \emph{Spitzer} IRS high-resolution spectrum of AU Mic (5-35$\mu$m) taken in 2005. That spectrum has unfortunately never been published, and the data reduction and especially background subtraction is not trivial, but a quick glance at that spectrum does hint at an excess starting at $\sim$18$\mu$m. Unfortunately, the IRS spectrum and the MIPS measurement are not compatible, and resolving the apparent contradiction is difficult.

There are more caveats to this.  
Looking at the images in the \emph{WISE} image archive, the centering of the source seems to be inaccurate. Because of this, are cautious about these measurements of AU Mic. AT Mic, another M4.5 member of the $\beta$ Pictoris moving group, detected with an excess of $\sim$15\% at 24$\mu$m using \emph{Spitzer} \citep{Plavchan2009}, is not detected as having excess using our technique. It is also not perfectly centered in \emph{WISE}. It has an absolute 3.4$\mu$m magnitude similar to that of AU Mic, pointing out again that absolute magnitudes are a poor proxy for stellar temperatures / spectral types of pre-main sequence M-dwarfs.

\section{Comparison to earlier studies}
\label{sec:comp}

Several studies have been performed on infrared excesses for main-sequence stars. We concentrate here on studies of 25$\mu$m and shorter-wavelength excess. The first major survey of this kind was conducted with IRAS. However, IRAS was on the whole not sufficiently sensitive for M-dwarfs \citep{Backman1993}. \citet{Su2006} searched $\sim$160 A-stars for excess with \emph{Spitzer}, yielding an excess rate of $32^{+5}_{-5}\%$. \citet{Moor2011} concentrated on F-type stars and observed 82 stars with \emph{Spitzer}, of which 15 show excess at 24$\mu$m ($18^{+5}_{-4}\%$), but in this case, the sample selection was biased toward stars that had shown hints for excess in other surveys. A more recent survey combining \emph{Spitzer} and \emph{WISE} data for a total sample of 263 F-stars detected excess in 19 sources, corresponding to $7.2^{+1.7}_{-1.5}\%$ (Mizusawa et al., 2012, submitted). \citet{Trilling2008} observed nearly 200 sun-like (FGK) stars and found that the incidence of debris disks detectable at 24$\mu$m is $4.2^{+2.0}_{-1.1}\%$. These authors also specifically targeted binary and multiple early-type (A3-F8) systems and found that the incidence of debris disks is comparable to or slightly higher than for single stars \citep{Trilling2007}. \citet{Carpenter2009} also targeted solar-type stars, concentrating on the evolution of detectable excesses with age. They found that the fraction of systems with detectable excesses ($>10.2\%$ above the photosphere) at 24$\mu$m drops from $15^{+3}_{-3}\%$ for ages $<300$ Myr to only $2.7^{+1.9}_{-1.3}\%$ for older ages. \citet{Bryden2009} used \emph{Spitzer} to determine whether the incidence of detectable excesses is different for stars known to harbor planets from radial-velocity searches. They found that any observed difference to stars not known to harbor planets is not statistically significant, but add to the general picture by contributing observations of 104 stars of spectral types F through M of which four ($2.7^{+1.6}_{-1.2}\%$) show detectable excesses at 24$\mu$m. None of their excess sources is an M-dwarf.

All these surveys targeted stars earlier than spectral type M. \citet{Plavchan2005} pointed out that there is a lack of debris disk detections for mature M-dwarfs and discussed possible reasons. \citet{Gautier2007} performed a \emph{Spitzer} search around nearby 62 M-dwarfs (some of their targets coincide with ours) and found no excesses at 24$\mu$m.

The study by \citet{Fujiwara2009} was performed using the data from the all-sky \emph{AKARI} satellite. Their detection threshold was significantly higher at $50\%$ above the photosphere (at 18$\mu$m), where they found significant excess around 14 main-sequence stars and derived an excess fraction of $1.5\%$. None of their excess stars is a K or M-dwarf.

\subsection{Dust at similar temperature}

As described above, we detected no significant excess that holds up to critical assessment in any of our stars. Because we studied 85 targets in the 3.4 and 12$\mu$m bands and 84 targets in the 22$\mu$m band, we derive an excess fraction (at the described levels) of $0^{+1.3}_{-0.0}\%$ amongst M-dwarfs. We did not consider the physically real excess around SCR 1845-6357 A because it stems not from a debris disk, but from a T dwarf binary. T dwarf binaries will not produce detectable amounts of excess around AFGK-stars, given current photometric precisions.

The discussion is somewhat complicated because for some of the targets, we have more than one target in the beam. Considering only targets with single stars in the beam (70 at 3.4 and 12$\mu$m, 69 at 22$\mu$m) yields an excess fraction of $0^{+1.6}_{-0.0}\%$, with some of them being part of wide-separation (distinguishable by \emph{WISE}) binaries. However, \citet{Trilling2007} showed that the excess fraction in binary systems is comparable to or slightly higher than that for single stars for a sample of A and F main-sequence stars. Furthermore, not all studies mentioned above considered only single stars. Because of this, we decided to directly compare our derived excess fraction of $0^{+1.3}_{-0.0}\%$ to the studies of AFGK and M-dwarfs performed in the literature.

Our derived detection rate for M-dwarfs is lower than for any other spectral class. This effect cannot simply be explained by the fact that our survey is less deep. While we reach only a median excess detection limit of $17.9\%$ in the $[3.4]$-$[22]$ color (to be compared with usually $\sim10-15\%$ for the \emph{Spitzer} surveys), our median excess detection limit in the $[12]$-$[22]$ color is $7.0\%$. In the most sensitive temperature range, the $[12]$-$[22]$ color usually gives the strictest limits (c.f. Figure \ref{fig:ProxCen}). We conclude that we should have been able to detect a comparable percentage of excess systems as would have been possible with a \emph{Spitzer} survey of the same targets.

Comparing our findings to the A-star survey of \citet{Su2006}, we can conclude that (randomly selected) M-dwarfs have a lower fraction of excess systems at a $\sim6\sigma$ level. Comparing our results to the F-star sample of Mizusawa et al. (2012, submitted), the difference of the combined M-dwarf surveys is significant on a $\sim3.6\sigma$ level. However, as shown by \citet{Carpenter2009}, the fraction of excess systems is strongly dependent on age. A- and F-stars are preferentially younger (because they live for a shorter time). Our sample of nearby M-dwarfs is likely to be sampled from the average population of M-dwarfs in our galaxy, we assume an average age of several Gyr (the star formation rate of the Milky Way peaks at ages $\sim$2-6 Gyr, see \citealp{Wyse2009}). The stellar age of the sample from \citet{Su2006} is on average only a few hundred Myr, with no star older than 1 Gyr. Mizusawa et al. (2012) provided no exact ages but estimated that their sample is probably on average $\sim300$ Myr. 
\citet{Carpenter2009} found that the excess fraction for FGK-stars drops to $3.6\%$ for ages 300-1000 Myr and $1.8\%$ for ages $\ge$1000 Myr. They also found that the upper envelope for 24$\mu$m excess at ages $\gtrsim500$ Myr is $\sim10\%$. If the same was true for our M-dwarf sample, of which the vast majority is expected to be older than 300 Myr and a majority to be older than 1000 Myr, this means three things: 1) The declining excess fraction with age can explain the scarcity of M-dwarf excess detections at 24$\mu$m and shorter wavelengths, 2) the excess fraction of M-dwarfs is not, on a 3$\sigma$ level, statistically different from the excess fraction of FGK-stars, and 3) we might be missing some debris disks very close to our detection threshold.

Our results agree with \citet{Gautier2007}, who did not find any excess around M-dwarfs. When combining the two samples for a total number of 112 M-dwarfs probed for excess at either 22 or 24$\mu$m and making the conservative assumption that $25\%$ of these M-dwarfs are in the 300-1000 Myr age range and the rest is $>1000$ Myr old, the excess fraction of the M-dwarf sample is different from the FGK sample on a $\sim1.1\sigma$ level, 
there is no clear evidence for a fundamentally different excess behavior of M-dwarfs at these wavelengths (corresponding to warm dust), but because of the general scarcity of M-dwarf excesses, more data are needed to clarify this question.

\subsection{Dust at similar radii}
\label{sect:radii}

In the last section, we have compared our 22$\mu$m (non)detections for M-dwarfs with studies of earlier-type stars at 24$\mu$m. This means, however, that we compared dust at very different radii. The orbital radius of dust of a given temperature, assuming blackbody grains, scales as $R_{dust} \sim L^{1/2}$. This means that while we probed dust at scales $\ll1$ AU for our M-dwarf sample, the same wavelength corresponds to dust at $\sim1$ AU for solar-type stars and $>1$ AU for A-stars. 

The discussion is additionally complicated because the dust behaves distinctly different from blackbody grains. For M-dwarfs, blowout due to radiation pressure is not relevant because of their low luminosities. This means that the dominant grain removal processes will be Poynting-Robertson (P-R) and stellar wind drag \citep{Plavchan2005, Saija2003}. Depending on the stellar wind from the M-dwarfs, the stellar wind drag can dominate. Since the stellar winds of M-dwarfs are not well known, their effects are difficult to quantify \citep{Plavchan2005} and the expected grain size distribution around M-dwarfs is unknown. On the other hand, grains smaller than the peak wavelength of their thermal emission will not behave like blackbody grains, but instead have higher effective temperatures than blackbody grains at the same location around the star. This means that we are in fact probing orbital radii larger than expected for blackbody grains with our observations (but likely still significantly smaller than the orbital radii we probe with observations at the same wavelengths around earlier-type stars).

Excess emission around stars of any type becomes rarer for shorter wavelengths. Excess at wavelengths shorter than 22 $\mu$m around main-sequence FGK-stars is very rare (see e.g. \citealp{Carpenter2009}). Since dust at similar radii would emit at shorter wavelengths for earlier-type stars, and since we were unable to establish a statistical difference in emission at the same wavelength, we conclude that we cannot establish a statistical difference for dust at similar radii either. We note, however, that the exact behavior of dust around M-dwarfs is complicated and potentially different from the evolution around earlier-type stars, and that a detailed analysis would require modeling the dust population.

Some detections of very hot dust very close to the star emitting in the $K_s$ band have been made through interferometry (e.g. \citealp{Absil2009, diFolco2007}). However, this dust is close to the sublimation limit and might represent a very different dust population than what we probed with our study. 

\subsection{Mass limits}
\label{sect:masses}

The mass limit for the dust is coupled to the total fractional excess (together with the albedo, this can be converted into the total optical depth around the star) as well as the distance from the star. The closer the dust is to the star, the lower the total dust mass.

As pointed out above, the exact location of the dust is difficult to calculate. However, we note that in general, we are more sensitive to small amounts of dust than comparable studies for earlier-type stars. \citet{Gautier2007} pointed out that their 24 $\mu$m observations of comparable depth led to upper limits for the dust mass as low as $10^{-9}M_{\oplus}$, although they did not state their assumptions. Because the location of blackbody dust grains scales as $R_{dust} \sim L_{\star}^{1/2}$, the dust mass detection limit scales as $M_{dust} \sim L_{\star}$, so the main reason we are able to place very strict limits on the dust masses is the low luminosities of the M-dwarfs.

Assuming blackbody grains of Bond albedo 0.3 and a single grain size of 1 $\mu$m with a density of 2.5 $g/cm^3$, we arrive at upper limits for the dust mass (3$\sigma$) of $1.7*10^{-9}M_{\oplus}$ for the 0.0035 $L_{\sun}$ Barnard's star and $2.5*10^{-9}M_{\oplus}$ for the 0.011 $L_{\sun}$ star GJ 526 for dust of $\sim$ 150K, and even stricter limits for dust at higher temperatures. We cannot make statements about dust at low temperatures, large amounts of which could be hidden from our survey at larger orbital radii. However, we are not aware of any detections of far-IR or sub-mm excesses for these objects.

\section{Implications}
\label{sect:SPHERE}

The dust evolution around M-dwarfs is still significantly less constrained than the evolution around earlier-type stars. The evolution of dust might follow a different path due to the lower luminosity combined with the low surface temperatures (and thus peak emission at longer wavelengths, longer than the smallest grain sizes). Blowout seems to be irrelevant for M-dwarfs. Theoretical models developed by \citet{Plavchan2005} imply that the fractional infrared luminosity $L_{IR}/L_{\star}$ might be independent of age for old M-dwarfs and there might be a fractional excess of $\sim9*10^{-7}$ for all M-dwarfs. It is important to point out that we would not be able to detect such an excess. We furthermore realize that an equal excess for \emph{all} stars in our sample would not be detected at all - because it would be masked by our fitting routine.

More investigations will be required to understand M-dwarf debris disks in more detail. While the \emph{Spitzer} 24$\mu$m channel is not available any longer and it will be quite a while until we can receive data from \emph{JWST}, there are other promising techniques available to find debris disks around M-dwarfs. For example, \emph{Herschel} has strong capabilities in the far-infrared, which allows one to study the frequency of colder dust. Studies at even longer wavelengths are possible (and have been conducted) in the sub-mm regime. To arrive at a complete picture of the M-dwarf debris disk population, both short- and long-wavelength studies will be required.

There are other interesting techniques on the horizon as well. Up to now, infrared excess studies have always been more sensitive to circumstellar dust than resolved imaging and therefore the amount of debris disks detected by their SED excess emission is much larger than the number of resolved debris disks. With the upcoming next-generation VLT instrument SPHERE and its high-contrast polarimetric imager called ZIMPOL, \citep{Schmid2006}, this could change. Owing to the use of the polarimetric differential imaging technique (e.g. \citealp{Quanz2011}), combined with an extreme adaptive optics system, ZIMPOL will have a very small inner working angle (IWA) of $\sim0.02\arcsec$. Because a dust ring this close to the star only occupies a small area on the sky, the surface brightness is high even when the total luminosity of the disk in scattered light is very low.

Assuming an unfavorable face-on viewing angle, blackbody grains, a thin dust ring compared to detector resolution, a geometric albedo of 0.3 and scattering albedo * polarization fraction = 0.1, we estimated the performance of SPHERE in detecting such rings. We converted these predictions into detection limits in terms of $F_{IR}/F_{\star}$ and plot these limits in figure \ref{fig:ProxCen}. We used preliminary performance estimates for ZIMPOL (achievable contrast of 7.8 / 8.1 / 9.5 ($mag/arcsec^2 - mag$) at separations of 0.02 / 0.2 / 1 $\arcsec$). SPHERE is expected to be more sensitive to warm dust than the analysis we present in this paper. The steep cutoff toward higher temperatures arises from the inner working angle of ZIMPOL. As we discussed in section \ref{sect:radii}, the real dust of a given temperature might be farther out than expected from a blackbody grain calculation because of the grain size distribution. Therefore, this might be able to probe warmer dust. Furthermore, this technique should work better (up to higher temperatures) for nearby sources. For example, for Barnard's star at 1.83pc, we can probe warmer dust than for GJ 526 at a distance of 5.41pc. SPHERE will not replace surveys in the mid- and far-infrared to sub-mm, but allow complementary observations of nearby sources, probing them for hot dust at small separations and attempting to resolve the dust spatially. It allows one to target individual stars to clarify uncertain detections or systems that are on the edge of having significant excess, such as the GJ 860 A/B system discussed in section \ref{sec:GJ860}. It will also be able to study known debris disks such as that of AU Mic or possibly AT Mic.

Together, all these techniques will hopefully enable us to understand not only the M-dwarf debris phenomenon, but debris disks in general in a more detailed way, including their frequency, their evolution, and their structure.

\section{Conclusions}

We performed a volume-limited search for mid-infrared excesses at 12 and 22$\mu$m using the RECONS database and the results of the \emph{2MASS} and \emph{WISE} all-sky surveys. Our main findings are:

\begin{enumerate}

\item
It is possible to fit the infrared colors of M-dwarfs with a polynomial using either the absolute magnitude in the first \emph{WISE} band (3.4$\mu$m) or the $V$-$K_s$ error as a proxy for stellar temperature. The resulting fit enabled us to search for excesses down to typically 14.9$\%$ for the $[3.4]$-$[12]$ color, 17.9$\%$ for the $[3.4]$-$[22]$ color and 7.0$\%$ for the $[12]$-$[22]$ color ($3\sigma$ median values).

\item
The quality of the \emph{WISE} data is excellent. The reduced $\chi^2$ goodness-of-fit values for the polynomial fits suggest that either the \emph{WISE} errors are overestimated or that there is a significant portion from a systematic calibration uncertainty in these errors, which would be corrected for by our fitting routine.

\item
No excess from a debris disk is found in any color within our detection limits. One physical excess is detected around SCR 1845-6357 A, but is shown to come from the T6 companion orbiting this late M8.5 dwarf. This proves the validity of our method. We also found a tentative excess around AU Mic at 22$\mu$m, but it is not part of our statistical sample.

\item
Compared to surveys for earlier-type stars, our derived excess fraction of $0^{+1.3}_{-0.0}\%$ is lower, but this is probably an age effect. The excess fraction in the mid-infrared (warm debris) is not significantly different from the excess fraction for old FGK dwarfs.

\item
Dust evolution around M-dwarfs is complicated and comparisons to earlier-type stars at either the same orbital distance of the dust or the same total dust mass are difficult because the primary clearing mechanisms for small dust are different. 
We note, however, that the limits on the total dust mass that can be derived from our survey are more stringent than the typical limits for earlier-type stars.

\item
The upcoming SPHERE/ZIMPOL instrument will search selected targets for reflected light from warm, close-in dust. This technique should be more sensitive for nearby bright stars than either \emph{WISE} or \emph{Spitzer} and could provide a mechanism to probe dust at very small angular separations.

\end{enumerate}

We conclude that \emph{WISE} is a great tool for studying the infrared properties of M-dwarfs and stars in general and will in the future provide a great source of information, just like IRAS did almost three decades ago. The detection limits achievable with \emph{WISE} are close to the limits achievable with pointed observations from \emph{Spitzer}.

In general, there is still a severe lack of knowledge about dust around M-dwarfs and more investigation is required both theoretically and observationally to understand the similarities and differences of debris dust evolution between M-dwarfs and earlier-type stars. While instruments like Herschel can help us to learn more about dust at far-infrared wavelengths, WISE and SPHERE/ZIMPOL can help by probing M-dwarfs for closer-in, warm dust rings.

\begin{acknowledgements}
This publication makes use of data products from the Two Micron All Sky Survey, which is a joint project of the University of Massachusetts and the Infrared Processing and Analysis Center / California Institute of Technology, funded by the National Aeronautics and Space Admi- nistration and the National Science Foundation.

This publication makes use of data products from the Wide-field Infrared Survey Explorer, which is a joint project of the University of California, Los Angeles, and the Jet Propulsion Laboratory / California Institute of Technology, funded by the National Aeronautics and Space Admnistration.

This research has made use of the SIMBAD database, operated at CDS, Strasbourg, France.

This research has made use of the RECONS database (www.recons.org).

We thank Jarron Leisenring for his help with the IRS spectrum of AU Mic.

Special thanks to Eric Mamajek for his valuable last-minute help with the spectral types of our sample stars.

We thank our unknown referee for his/her very detailed and very helpful referee report, which allowed us to improve the quality of this paper.
\end{acknowledgements}

\nocite{*}
\bibliographystyle{aa.bst}
\bibliography{mybib.bib}

\appendix
\section{Details on the fitting routine}
For the determination of a fit curve to the color-magnitude data required for the subsequent search for excess sources, we used a polynomial fitting routine. We chose a polynomial degree $n$ and then proceeded as follows:

\begin{enumerate}
\item Select only the single stars from the sample of all stars
\item Perform a least-squares fit for polynomial of degree $n$
\item Calculate $\chi^2$ goodness-of-fit
\item Calculate $\sigma' = \frac{\sigma}{\sqrt{\chi^2}}$ for all stars
\item Remove stars with $|\sigma'| > 3$ from the sample
\item Repeat from 2) until converged
\end{enumerate}

We did this for polynomials of increasing degree, starting with $n=1$. For all colors, the quality of the fit does not increase significantly beyond $n=4$, therefore we chose to use polynomials of fourth degree in all cases.

\begin{figure}
\includegraphics[width=8.8cm]{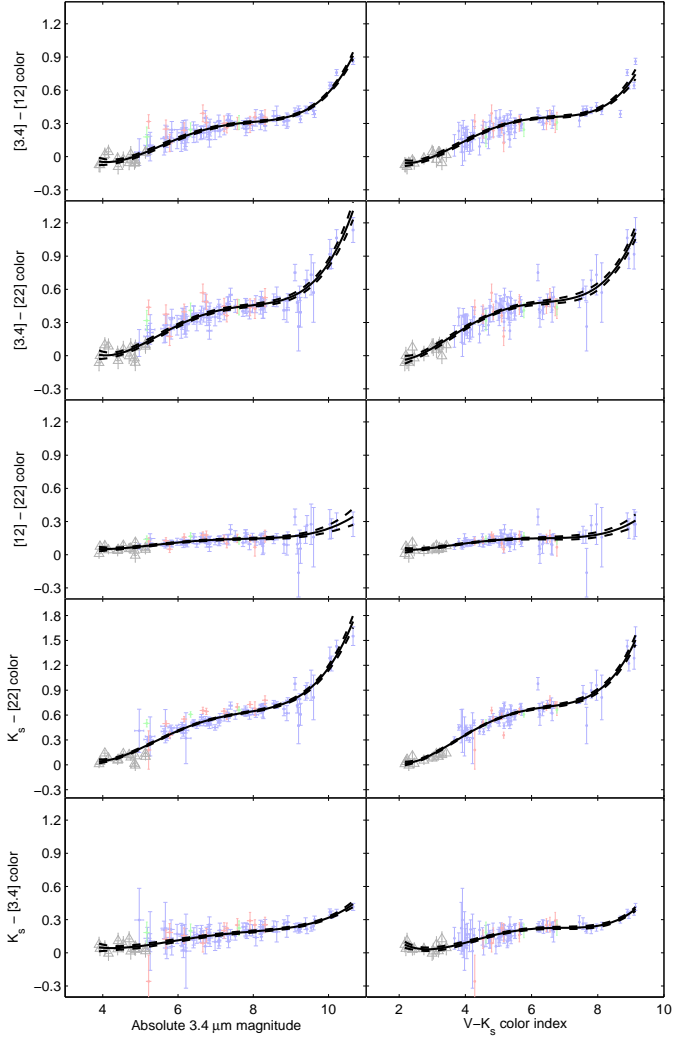}
\caption{Obtained fits and error estimates for all three colors and additionally the $K_s-[22]$ and $K_s-[3.4]$ colors against absolute magnitude in the first WISE band and against $V$-$K_s$ color.}
\label{fig:FitErrors_All}
\end{figure}

\begin{figure}
\includegraphics[width=9cm]{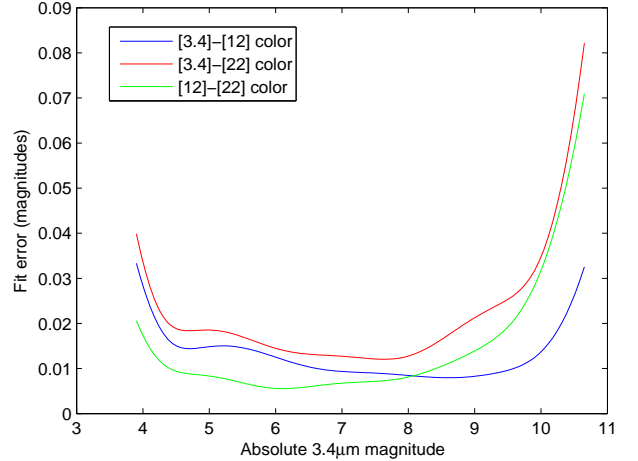}
\caption{Inferred errors for our fits. As can be seen, the fit is very well constrained for intermediate M-dwarfs, but less constrained at both ends of the fit. At the high-luminosity end, we mitigate this problem by including several K-dwarfs.}
\label{fig:FitErrorsW1}
\end{figure}

\begin{table*}[ht!]
\caption{Polynomial fits}
\label{table:fits}    
\centering
\begin{tabular}{cccccccc}
\hline\hline\noalign{\smallskip}
{Color} & {Proxy}&    {$a_0$}& {$a_1$}& {$a_2$}& {$a_3$}& {$a_4$}& {$\chi^2$}\\
\noalign{\smallskip}
\hline
\noalign{\smallskip}
{$[3.4]-[12]$}&        {$M_{3.4}$}&   {5.6839e+00}&  {-3.9338e+00}&   {9.5780e-01}&  {-9.7494e-02}&   {3.5935e-03}&{0.6943}\\
{$[3.4]-[22]$}&        {$M_{3.4}$}&   {8.2659e+00}&  {-5.6808e+00}&   {1.3899e+00}&  {-1.4269e-01}&   {5.3063e-03}&{0.6054}\\
{$[12]-[22]$}&        {$M_{3.4}$}&   {1.7381e+00}&  {-1.1721e+00}&   {2.8991e-01}&  {-3.0074e-02}&   {1.1295e-03}&{0.5057}\\
{$K_s-[22]$}&       {$M_{3.4}$}&   {6.6826e+00}&  {-4.7923e+00}&   {1.2182e+00}&  {-1.2829e-01}&   {4.8876e-03}&{1.6584}\\   
{$K_s-[3.4]$}&       {$M_{3.4}$}&   {2.2181e+00}&  {-1.4508e+00}&   {3.4410e-01}&  {-3.4285e-02}&   {1.2482e-03}&{0.5662}\\
{$[3.4]-[12]$}&        {$V$-$K_s$}&   {5.5283e-01}&  {-7.6010e-01}&   {2.9965e-01}&  {-4.1715e-02}&   {1.9851e-03}&{1.0173}\\
{$[3.4]-[22]$}&        {$V$-$K_s$}&   {7.3936e-01}&  {-1.0151e+00}&   {4.2361e-01}&  {-6.2772e-02}&   {3.1802e-03}&{0.7858}\\
{$[12]-[22]$}&        {$V$-$K_s$}&     {2.7830e-01}&  {-2.9003e-01}&   {1.1597e-01}&  {-1.7087e-02}&   {8.6622e-04}&{0.5322}\\
{$K_s-[22]$}&       {$V$-$K_s$}&   {1.2611e+00}&   {-1.4911e+00}&   {5.8420e-01}&  {-8.3413e-02}&   {4.1207e-03}&{2.4038}\\
{$K_s-[3.4]$}&       {$V$-$K_s$}&   {1.2223e+00}&  {-1.1065e+00}&   {3.5304e-01}&  {-4.4812e-02}&   {2.0075e-03}&{0.4231}\\
\noalign{\smallskip}
\hline
\end{tabular}
\tablefoot{Parameters of our estimated polynomial fits for the different colors and the different stellar temperature proxies (3.4$\mu$m absolute magnitude and $V$-$K_s$ color). A $\chi^2$ goodness-of-fit value is given for reference. Description of the parameters in the text.}
\end{table*}

From the $\chi^2$ value of the fit, it is possible to estimate the amount of intrinsic scatter in the stellar colors, assuming the error estimates for WISE are correct. As seen before, the WISE errors are likely to be overestimated because the $\chi^2$ values for all three fits are lower than one. We conclude that the intrinsic scatter is small and the actual deviations from the fit are dominated by measurement errors. This is true for all WISE colors ($[3.4]$-$[12]$, $[3.4]$-$[22]$ and $[12]$-$[22]$). The $K_s-[22]$, which we also investigated and show here for reference, has a higher $\chi^2$ value, and intrinsic scatter (induced by scatter in the $K_s$ color) might play a more substantial role here. $K_s-[3.4]$ is very good again, with a low $\chi^2$ value. We show it here for reference as well because it bridges the \emph{2MASS} longest and \emph{WISE} shortest wavelengths for M-dwarfs, which we think could be useful.

While it is possible to obtain uncertainties on the derived parameters of a polynomial fit, this is of limited helpfulness because the individual parameters are highly correlated. To estimate the error on our fit, we again took a data-driven approach. We used the unmodified $\sigma$ errors from WISE (not $\sigma'$) and produced 1000 data sets by randomly adding Gaussian errors with a standard deviation of $\sigma_i$ to each measurement. We fit these 1000 data sets using our fitting routine. We then calculated the uncertainty of our fit at every point by calculating the standard deviation of these 1000 obtained polynomial fits.

Figure \ref{fig:FitErrors_All} summarizes the fits and their upper and lower limits (dashed lines). Because the errors are small and hard to see in this representation, Figure \ref{fig:FitErrorsW1} shows the error of each fit as a function of the absolute magnitude in the first WISE band. As can be seen, the errors are generally small, mostly below 0.02 magnitudes, and thus much smaller than the measurement errors from the individual stars, which is why we did not provide an in-depth discussion of fitting errors in our analysis section (the error is not dominated by the fitting error). At both ends of the fit, the fitting errors naturally increase because the fact that the fit is less constrained at the ends. For the high-luminosity end of the M-dwarfs, this effect can be mitigated by including some K dwarfs, as we have done. We have good reasons to believe that our fundamental assumption that the colors can be fitted by a (fourth-order) polynomial holds reasonably well in the stellar regime (namely, the $\chi^2$ values of our fits are very good).

We summarize the parameters of our obtained fourth-order polynomials in table \ref{table:fits}. The expected color can then be calculated as $$color(p) = a_0 + a_1 p + a_2 p^2 + a_3 p^3 + a_4 p^4,$$ with p being the proxy for the stellar temperature (either absolute 3.4$\mu$m magnitude or $V$-$K_s$ color). In addition to the three WISE bands, we also provide fits for the $K_s-[22]$ color here. 
The $K_s$-$[22]$ color can also be used for \emph{Spitzer} data using the small offset of 0.083 magnitudes determined in Fig. \ref{fig:SpitzervsWISE}.

\end{document}